\pdfoutput=1
\documentclass[a4paper,12pt,twoside,openright]{report}

\def\authorname{Philippos Papaphilippou\xspace}
\def\authorcollege{Wolfson College\xspace}
\def\authoremail{Philippos.Papaphilippou@cl.cam.ac.uk}
\def\dissertationtitle{Performance tuning for deep learning on a many-core processor}
\def\wordcount{14,327}

\usepackage{epsfig,graphicx,parskip,setspace,tabularx,xspace,soul} 
\usepackage[hidelinks]{hyperref}
\newcolumntype{P}[1]{>{\centering\arraybackslash}p{#1}}

\begin{document}

\pagestyle{empty}
\singlespacing
\begin{titlepage} 

\begin{center}
\noindent
\huge
\dissertationtitle \\
\vspace*{\stretch{1}}
\end{center}

\begin{center}
\noindent
\huge
\authorname \\
\Large
\authorcollege      \\[24pt]
\includegraphics{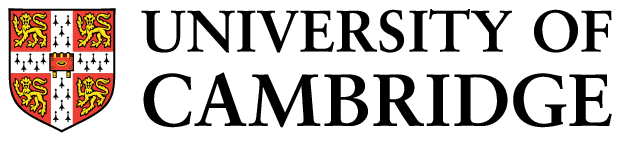}
\end{center}

\vspace{24pt} 

\begin{center}
\noindent
\large
{\it A dissertation submitted to the University of Cambridge \\ 
in partial fulfilment of the requirements for the degree of \\ 
Master of Philosophy in Advanced Computer Science} 
\vspace*{\stretch{1}}
\end{center}

\begin{center}
\noindent
University of Cambridge \\
Computer Laboratory     \\
William Gates Building  \\
15 JJ Thomson Avenue    \\
Cambridge CB3 0FD       \\
{\sc United Kingdom}    \\
\end{center}

\begin{center}
\noindent
Email: \authoremail \\
\end{center}

\begin{center}
\noindent
June 18, 2017
\end{center}

\end{titlepage} 

\newpage
\vspace*{\fill}

\onehalfspacing
\newpage
{\Huge \bf Declaration}

\vspace{24pt} 

I \authorname of \authorcollege, being a candidate for the M.Phil in
Advanced Computer Science, hereby declare that this report and the
work described in it are my own work, unaided except as may be
specified below, and that the report does not contain material that
has already been used to any substantial extent for a comparable
purpose.

\vspace{24pt}
Total word count: \wordcount

\vspace{60pt}
\textbf{Signed}: 

\vspace{12pt}
\textbf{Date}:

\vfill

This dissertation is copyright \copyright 2017 \authorname. 
\\
All trademarks used in this dissertation are hereby acknowledged.

\newpage
\vspace*{\fill}

\newpage
{\Huge \bf Acknowledgements}
\vspace{24pt}

I would like to thank Dr. Daniel Bates, Alex Chadwick and my supervisor Dr. Robert Mullins for their insightful feedback and dedication during our weekly meetings. Dr. Daniel Bates and Alex Chadwick have also supported my project by implementing my selected variations of the convolution algorithm on Loki. I would also like to thank my family for their support during all my university years.

\newpage
\vspace*{\fill}

\singlespacing
\newpage
{\Huge \bf Abstract}
\vspace{24pt} 

Convolutional neural networks (CNNs) are becoming very successful and popular for a variety of applications. The Loki many-core processor architecture is very promising for achieving specialised hardware performance and efficiency while being a general purpose solution. Loki combines many simple cores with increased control for the programmer. This freedom can be exploited to produce much more efficient code than in conventional multiprocessors but it also creates a very big design space for possible optimisations. In this project, I explore possible optimisations for a CNN application, their portability on different Loki-specific configurations, convolution parameters and inputs. Finally, I investigate the potential for adaptive algorithms for further performance increase.

\newpage
\vspace*{\fill}

\pagenumbering{roman}
\setcounter{page}{0}
\pagestyle{plain}
\tableofcontents
\listoffigures
\listoftables

\onehalfspacing


\chapter{Introduction}
\pagenumbering{arabic} 
\setcounter{page}{1} 

Convolutional neural networks are becoming very successful and popular for image recognition and speech recognition tasks \cite{6}. However, they require significant computational power and conventional architectures are proven to underperform considerably in comparison with specialized hardware. Specialised hardware, such as with the use of FPGAs, CGRAs and ASICs, introduce many challenges at different stages of development and maintenance \cite{7,8,9}. While specialised hardware could provide the optimal solution in terms of power efficiency and performance \cite{6}, GPUs and other general purpose platforms are a practical alternative because they offer reprogrammability while eliminating the challenges of specialised hardware.

One such general purpose platform that is promising for neural network applications is the Loki architecture \cite{1}. Loki is a many-core processor that utilises a simple interconnection network design to maintain high connectivity and throughput between 16 tiles of 8 cores each while keeping the power utilisation low. Loki also offers the functionality to swap tiles for L2 cache size programmatically. This option alongside other optimisation options, such as different algorithmic approaches, resource allocation schemes or memory access patterns, creates a design space for exploration for optimising this application for the target architecture.

In this project I explore the potential of these design options, possible optimal compile-time decisions and the benefit of optimisation during runtime. The latter may be useful in our application because the input data can vary substantially and a static configuration might not perform well in all cases. 

The Loki many-core processor will be a very capable chip and a very interesting object of study due to the flexibility it offers and probably a very high throughput for certain applications. The exhaustive analysis of a related optimisation design space in this project shows the dimensionality of the problem and also some smart approaches to perform quicker exploration for Loki and also for generalizing the findings for different micro-architectural and implementation specifications.

In Chapter 2, I provide background information related on the architecture and the problem. In Chapter 3, I demonstrate a set of trivial and non-trivial optimisations and discuss possible challenges for exploring them or applying them to Loki software implementations. In Chapter 4, I present my loop interchange analysis where I search for the top performing nested loop permutations. In Chapter 5, there is further analysis of the loop permutation results. In Chapter 6, I get more realistic performance results and finally in Chapter 7 there is a summary along with a list of possible future work.

\chapter{Background Information} 

In this chapter I present the basic principles behind the architecture, the convolutional neural networks and the experimental methodology. I also explain the available high-level parameters that can affect the overall performance. 

\section {The Loki architecture}
	
Loki is a general purpose many-core architecture architecture that aims to provide the flexibility found in CGRAs for better performance and power efficiency for specialization purposes. It consists of a number of homogeneous tiles of 8 cores and 8 memory banks each. Each memory bank is an 8-Kbyte SRAM that is connected to every core in the tile through a crossbar. The design decision for homogeneity was preferred in order to provide fault tolerance capabilities, modularity in design and verification and scaling. One of the advantages of Loki is that its Instruction Set Architecture provides more control to the hardware, such as the ability to disable tiles for a unified L2 cache or more direct inter-core communication with packets.

The components are low-end in comparison with modern multi-cores. For example the total size of memory that is used for L1 and L2 purposes is 1 MByte across the whole 128-core chip configuration, when the total number of tiles is 16. In addition, there are not any cache coherence hardware mechanisms, but the related commands are exposed for the user to implement coherency in software. This design decision lowers the complexity and power usage of the components and but also enables the programmer or compiler to exploit more performance out of specialised applications. The many-core topology favours applications with a high degree of explicit parallelism but it also provides better performance than reconfigurable architectures for control-intensive software. It supports many parallel programming paradigms such as equivalent functionalities for SIMD, fine-grain dataflow, task-level pipelines, Instruction-Level parallelism and others \cite{2}.

In Figure 2.1 we can see a graphical representation of the Loki processor. Regarding intra-chip communication, each tile is connected to an on-chip interconnection network, consisting of 3 separate networks, a request, a response and a reply network for deadlock avoidance. Both data and instructions are transferred as packets. Each core communicates with main memory through channels. Their mapping is decided by the the source code and the assignments are stored in the channel map table (CMT). The cores and memory banks inside a tile also communicate with each other by using crossbars. The memory banks of a tile are also interconnected with a ring network for supporting cache-related functionalities. The cores are also interconnected and this is achieved with multicast buses.

\begin{figure}[h]
\centering
 \includegraphics[width=0.7\textwidth]{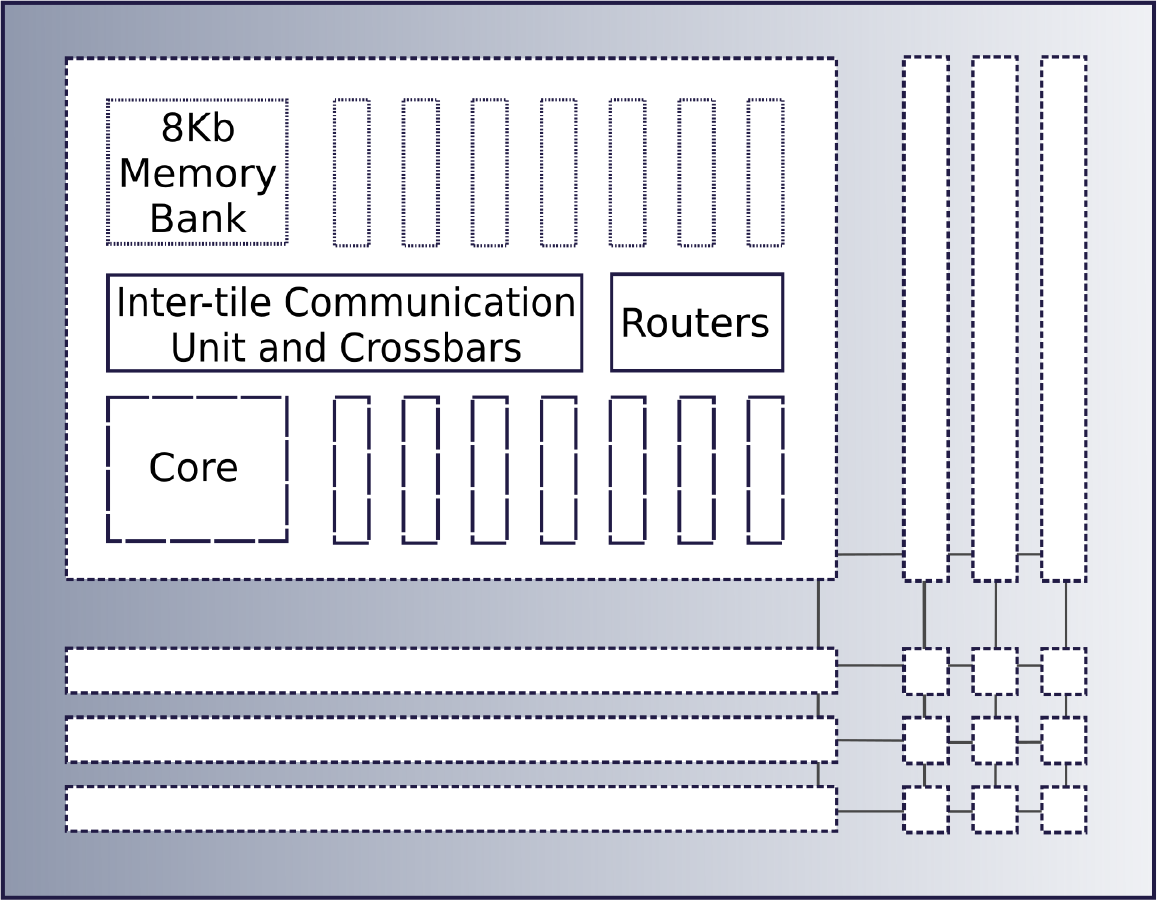}
\caption{Loki architecture diagram for 4x4 tiles.}
\end{figure}

The combination of modularity and flexibility here is both a curse and a blessing. On the one hand the design process and verification became less complex as modular designs easier to validate and scale. Possibly, it will also be future-proof, as the less hardware specialisation leaves more room for wider adoption and future software optimisation. On the other hand it lacks high performance features, such as hardware floating point support, multiple levels of set-associative caches, SMP cores, a big Last-Level Cache (LLC), specialised instructions for accelerating complex operations, hardware coherency mechanisms and many others. It also lacks more complex mechanisms that are applied in modern multi-core processors for increasing the Instruction-Level Parallelism (ILP), such as wider pipelines, data prefetchers, sophisticated branch predictors, as well as a modern replacement policy for the LLC cache, instead of random replacement. 

While the absence of many of those mechanisms could be compensated by more intelligent software, especially for applications that are parallelisable at a high degree and have certain workload characteristics, the biggest problem is the huge design space it provides for software optimisation. The Instruction Set Architecture gives so much control to the software that development for this platform would probably be effective only in specialized applications, where the programmer has a deep understanding of the architecture. 

The current state of the compiler for Loki requires human intervention for the production of fast code. Ideally, the compiler will eventually become mature enough to apply a variety of optimisations such as automatic loop vectorisation with SIMD equivalent routines. Even then, the design space for optimisations would still be very big, as the architecture offers much greater amounts of additional functionality and freedom than regular multi-processors.

One other way Loki is able to compensate for the lack of powerful computing and local storage resources, it supports virtual architectures \cite{10}. The different pipeline stages and resources of each core can be used remotely and as a result this freedom can be used for emulating fewer more-complex cores and custom memory hierarchies.

The parameters that I will explore that are related to the architecture configuration and try to optimise are the \ul{number of tiles working as computation units} and the \ul{number of tiles working as a unified L2 cache}. I will also explore some task-specialization for the cores of each tile by evaluating some current Loki implementations of the convolutional neural network application. One of the goals of this project is to decide whether these options can take optimal values for all inputs or if there is a need for changing these values dynamically.   

\section{Convolutional Neural Networks}

Convolutional Neural Networks (CNNs) are a type of Artificial Neural Networks (ANNs) used for Deep Learning that uses convolution operations in its layers. They are very similar to the more classic Multi-Layer Perception (MLP) ANNs with respect to the feed-forward data flow and the multiple inner layers of neurons. 

In general, the difference between CNNs and MLPs is that CNN applies convolution between the majority of the first layers among transformations and sampling techniques, such as max-pooling \cite{6}. At the beginning there is a much bigger number of input neurons for inputs such as color images (RGB arrays) that it would be computationally very expensive to train the network if it was a fully connected MLP. Then the data goes through a number of those operations and ends up in a more compact form in the last layers which are fully-connected as with the MLP ANNs.

A convolutional layer applies a series of two-dimensional filters or kernels to multiple same-sized input 2-dimensional monochrome images using convolution \cite{6}. This can be computed using direct convolution which uses a sliding window to calculate the dot product. There are also alternative ways to calculate convolution, such as FFT Convolution, which is faster for bigger kernels \cite{16}. The convolution happens between the convolution part of the network as noted in Figure 2.2, where a Multi-Layer Perceptron is compared with a simplified form of a Convolutional Neural Network architecture.

\begin{figure}[h]
\centering
\includegraphics[trim=2cm 0 0 0,width=1.08\textwidth]{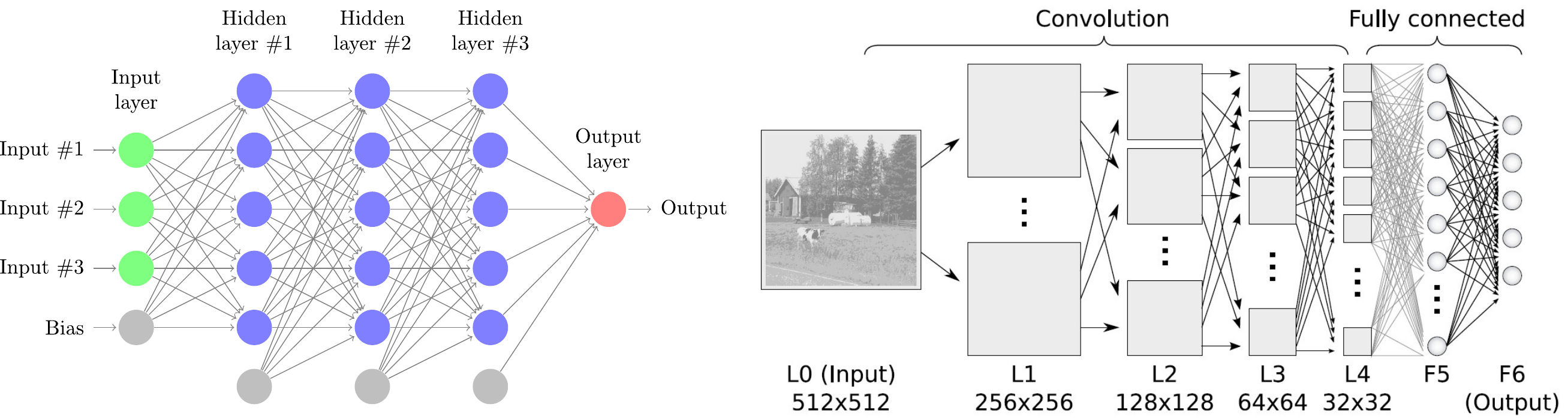}
\caption{Comparison of an MLP (left) and a convolutional neural network (right, source: \href{http://www.ais.uni-bonn.de/deep_learning}{University of Bonn, Autonomous Intelligent Systems})}
\end{figure}

Regarding the main computation, we note that the convolution of one layer takes a number of input images (number of input channels) and a number of kernels (equal to the number of the output channels) and performs convolution for every pair of the two sets and produces a 3D array that will be the input of the next layer. The number of iterations that direct convolution operation makes is equal to the input image area multiplied by the kernel area. This results in a total number of iterations equal to the product of the number of input channels, the number of \ul{output channels}, the \ul{width of the image}, the \ul{height of the image}, the \ul{width of the kernels} and the \ul{height of the kernels}. In a later chapter we will see an implementation consisting of six nested loops that performs convolution. All of these parameters will be taken into consideration to measure how performance is affected under different conditions.

The architectural decisions for the simpler Multi-Layer Perceptron Networks are the number of hidden layers, the number of neurons in each layer, the transfer functions between layers and the input functions. There might be several problems with explicitly constructed Neural Network Architectures \cite{11} such as poor performance after evaluating with k-Folds Cross Validation, long training times, or even incapability of converging to a specific amount of error. There are also parameters and optimisations related the the training algorithms. The optimisation of a Multilayer Perceptron to solve a specified problem is still an unsolved task \cite{11} and it is very desirable to build a constructive Neural Network Learning algorithm \cite{12} that would decide the optimal topology algorithmically. 

In the CNN case the design space is wider as many different types of layers with their own customisation option have been introduced that could be used in different arrangements when building the neural network topology. However, there have been studies that suggested certain architectures such as a famous CNN by Krizhevsky et al. \cite{13}, SqueezeNet \cite{14}, GoogLeNet and AlexNet \cite{15}. All these examples are used in variety of applications and are proven to perform relatively well especially in photograph classification \cite{15}.

The important common characteristic of Convolutional Neural Networks is that they are computationally expensive and the basic building block is convolution. Neural networks are generally known to be more expensive than other machine learning algorithms because they make less statistical assumptions on the input data. This is the reason that some of the algorithm design choices were already selected to increase performance. One example of this is the use of Rectified Linear Units (ReLU) which is a linear transfer function that is currently more favourable to the classic much more CPU-time-consuming tanh or sigmoid transfer functions \cite{13}.

This project will not focus on the neural network architecture design aspects. The goal will be to fine-tune the execution of the whole family of CNN algorithms by examining and optimising the most demanding workload of their execution, which is the convolution. The target architecture is attractive for this kind of workload as it is highly parallel. In addition, there is a variety of different amounts of data localities in convolution and they might be more easily exploited for better performance using Loki’s low cost communication features.

\section{Experimental Methodology}

There are two main simulation frameworks that I use for the analyses. First, I have developed a custom cache simulator with that is a fast functional simulator based on binary instrumentation that also makes rough performance predictions. Then I use lokisim, which is the official performance simulator of the Loki team, for validating the optimisations or monitoring bottlenecks.

The two simulators of the micro-architecture offer results based on two different levels of abstraction. The more detailed a simulator is, the more accurate the results will be. Additionally, the lower abstraction also increases time significantly as it models more aspects of the simulated microprocessor. Since the design space of exploration can be very large, a general methodology that is widely used is to make the exploration exhaustively using simpler simulations and then validate the most promising decisions in a more detailed simulation environment.  

\subsection{Cache Simulator}

In order to evaluate a big number of optimisation variations on many input and hardware configurations efficiently, I developed an Intel Pin \cite{17} tool based on pinatrace.cpp. It is a many-level cache simulator that has a form of a Pin tool with similar configuration to Loki’s memory hierarchy’s characteristics for faster design space exploration. Pinatrace.cpp ia a Pin tool that produces a stream of memory accesses and aims to be independent to the underlying architecture \cite{18}. This is also a major advantage because the instrumentation is done on normal linux binaries, which are less expensive to produce than Loki binaries which require human intervention in the code for efficient usage of the resources. 

Originally, the pinatrace.cpp Pin tool produced a multiple-gigabyte address stream file by instrumenting a binary that runs on the host machine architecture (x86). This practice is itself time consuming, as I/O operations are very expensive and also impractical when the number of combinations is that large. One solution would be to pipe the address stream from the Pin tool’s stdandard output to a separate cache simulator but after experimental evaluation, I found that it was around 40 times slower than the embedded simulator implementation. My current Pin tool produces the results for a single run in a couple of seconds in a summary form. Similar practices of summarised reports are well known for efficient probing/profiling results in many applications, such as the aggregation functions of DTrace \cite{19}, the Intel Pin example tools, as well as in hardware applications \cite{20}.

Table 2.1 summarises the initial cache simulator parameters to represent Loki’s design. In later stages the winner optimisation schemes are tested in lokisim, which has a lower level of abstraction and can give a more representative insight for overheads, such as the interconnection network bandwidth limit. One difference of the initial cache simulator configuration is that the L1 cache is shared between instruction and data blocks, while Loki has a separate small memory for instruction accesses. This configuration is used for simulations using up to 8 threads inside a single tile and with L2 cache equivalent to 8 tiles.

\begin{table}[h]
\centering

 \begin{tabular}{||c||P{1.6cm}|p{2.1cm}|P{1.8cm}|P{1.3cm}|P{1.5cm}|c||} 
 \hline
 Memory	level & Access	Latency & Size	(KBytes) & Block size	(Bytes) & Associ- ativity & Repl.	policy & Scope \\ [0.5ex] 
 \hline\hline
L1 cache & 3 cycles & 64 (1 tile) & 32 & 1 & - & Shared \\ 
 \hline
L2 cache&10 cycles&512 (8 tiles)&32&8&Random&Shared\\
 \hline
  Main memory &30 cycles&-&-&-&-&- \\ [1ex] 
 \hline
\end{tabular}
\caption{Cache simulator parameters to model a Loki design.}
\end{table}

The way with which the cache simulator provides performance estimates is simplistic. The total cycles are calculated by adding one cycle for each of the non-memory instructions and the number of hits in each memory level (L1, L2 and main memory) multiplied by their respective access latency. This high level of abstraction provides a very fast simulation infrastructure that is comparable to an off-the-shelf micro-architectural system simulator, such as MARSSx86, for the equivalent simplified model parameters to model Loki’s cache hierarchy.

\begin{figure}[h]
\centering
 \includegraphics[trim=1.2cm 0 0 0,width=1\textwidth]{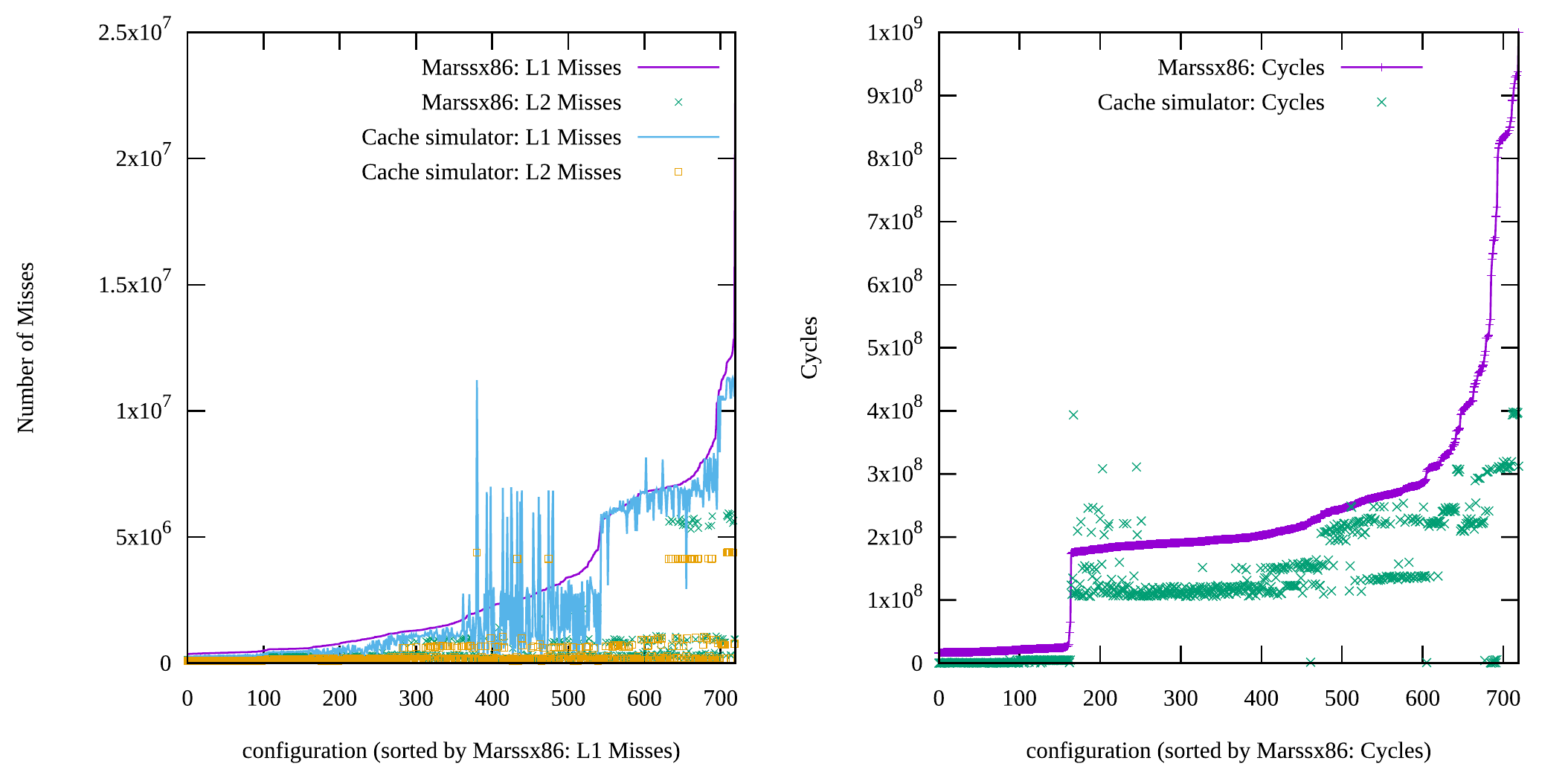}
\caption{Comparison of equivalent performance metrics from MARSSx86 and the custom cache simulator.}
\end{figure}

In Figure 2.3 we can see a rough comparison of different inputs of a single-thread application performance under MARSSx86 and my custom cache simulator. In order to model a very simple CPU core inside MARSSx86 I have set both the issue width and the dispatch queue size to be equal to 1. One important observation is that while the results can be noisy, the good configurations section at around the configuration 150 in Figure 2.3 (right) is correctly predicted by the cache simulator.

I have also created some other versions of the Pin tool that have different features. The additional functionality is the option to stop after a specific number of instructions, a multi-core/multi-tile version that is more related to Loki and the option to change the cache sizes from the command line arguments. I have also implemented Belady’s OPT optimal replacement policy \cite{27} as an option for the cache block replacement policy to replace the default random policy, for bottleneck analysis.

\subsection{Lokisim}

Lokisim is a high-level simulator for the Loki architecture built by the Loki team. It was written in SystemC and aims to be a fast alternative to a cycle-accurate verilog implementation. It supports a variety of simulation options including modeling a chip outside the design specifications as well as simulation-related options. The micro-architectural options are the number of total tiles, number of cores and memory banks per tile, the size of memory banks the size of instruction cache and many more. Some simulation options include the functionality to stop the simulation after a specific number of cycles and some unrealistic features such as zeroing the memory access latency for identifying bottlenecks. It also provides detailed statistics such as the number and type of memory accesses,  all operations usage breakdown for each core of each tile, the bandwidth usage, the average Instruction per Cycles and also the average link contention for inter-tile communication.

\chapter{Algorithmic Optimisations}

In this chapter I describe a set of algorithmic decisions or optimisations that can be applied in the main convolution loop to increase performance. Some of them create a design space whose exploration would aim to increase the data localities and data reuse, the elimination of redundant or unnecessary computations, as well as Loki-related application for more efficient utilisation of the chip’s resources. 

For some of the optimisations it might be difficult to make generalizations due to the high number of possible input parameters that change the workload characteristics and also because the number of the combinations of the decisions here is also very high even with a single specified input. There are separate chapters for analyses on the loop interchange analysis  and the tile vs L2 case, which is more specific to Loki. 

One of the scopes of the analysis is to decide whether some of these optimisations could be statically set and perform well in a variety of different input configuration parameters on run-time conditions or there would be benefit to have dynamic optimisations that adapt to the workload. 

\section{Basic optimisations}

In this section I present the basic steps to eliminate unnecessary multiplications in the most time consuming part of the convolution. These optimisations might be trivial but they are presented for better understanding of the algorithm and also they contribute to the methodology as we will later examine realistic access patterns. First, we transform the data from multiple dimension array representations to linear memory to better identify data localities and exploit optimisation opportunities. 

For example, 2D data of the form a[x][y] would be transformed in a[x*\texttt{<} size of Y dimension\texttt{>} +y]. An example for 3D data would be the transformation from a[x][y][z] to a[x * \texttt{<}size of Y dimension\texttt{>} * \texttt{<}size of Z dimension\texttt{>}  + y * \texttt{<}size of Z dimension\texttt{>} + z]. This is actually how basic array data types in common programming languages are represented in memory, such as with C/C++. Therefore, by this transformation we remove a layer of abstraction to allow manual optimisations. In Figure 3.1 we observe the differences after the transformation.

\begin{figure}[h]
\centering
 \includegraphics[width=1\textwidth]{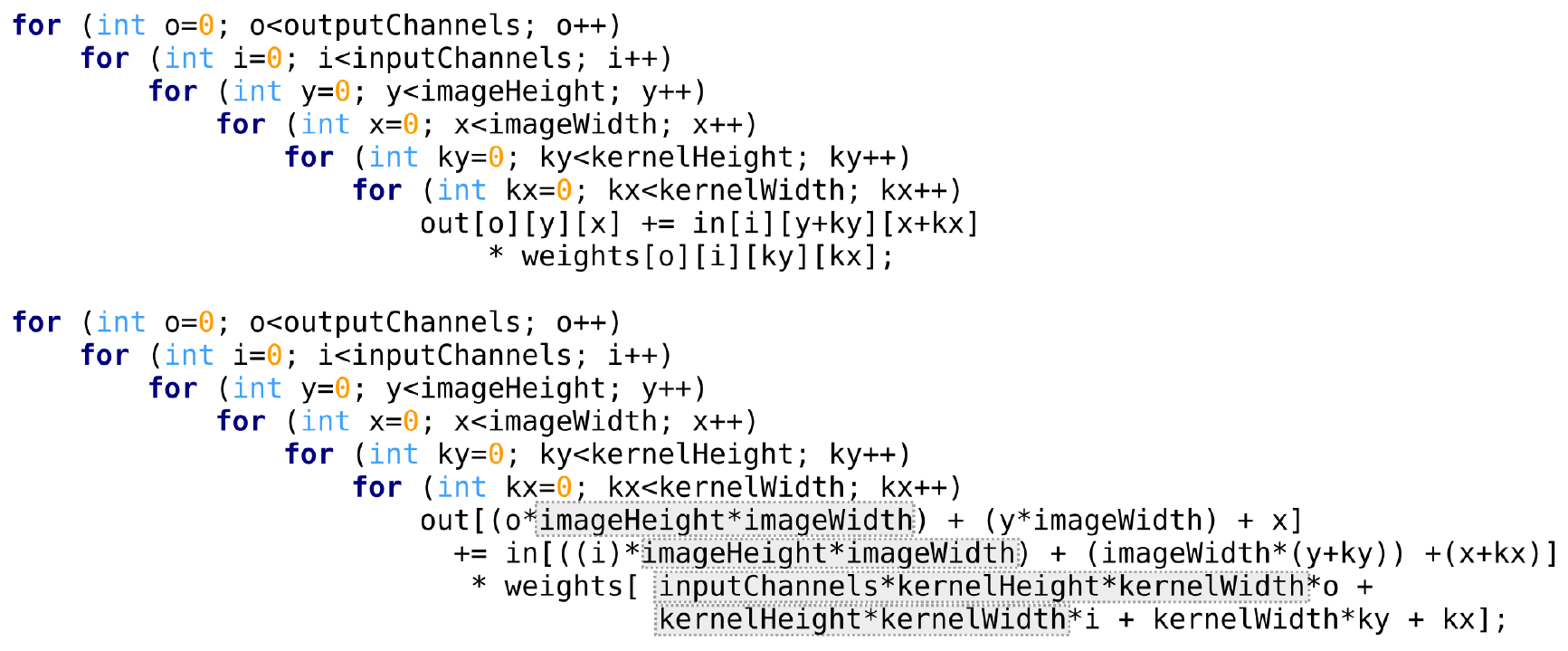}
\caption{The main nested loop of convolution (Top) and the implementation with one-dimensional arrays (Bottom). The gray areas highlight unnecessary multiplications. }
\end{figure}

As we notice, the shaded multiplications, such as the product of the image height and image width, have the same result in all iterations. Therefore, we can replace all the shaded portions with pre-calculated immediate values. This reduces the number of unnecessary frequent calculations. 

Another observation is that even after the replacement of those multiplications there still remain some other multiplications, that could be computed earlier, such as y*imageWidth which could be determined after the 3rd nested loop in the figure, which increments y. Instead of performing those multiplications earlier, it would be even better to perform additions on respective sums, as addition is less expensive than multiplication in general. In Figure 3.2, we observe the resulting main loop code, where each for structure appears as a building block and the arrows point to the respective closing section.

\begin{figure}[h]
\centering
 \includegraphics[trim=0.85cm 0 0 0,width=1.1\textwidth]{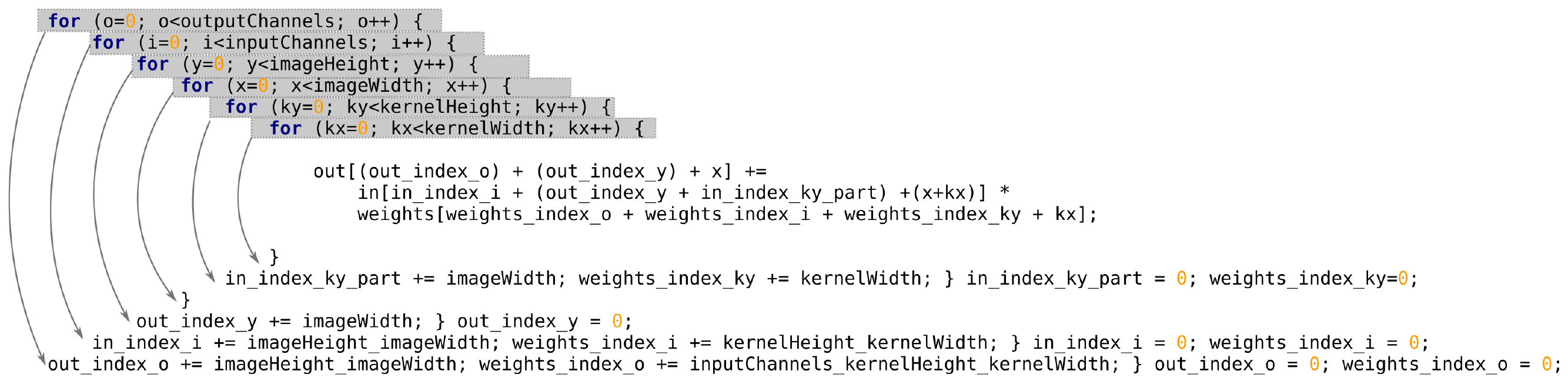}
\caption{Optimised main loop with structure to generate permutations easily.}
\end{figure}

We note that there is only one multiplication left, which is necessary for the convolution computation. There is one more similar optimisation; the values of o, i and y are not used internally, and therefore we could iterate over their dependent values instead of them. However this might have a small impact in performance and irregular for loop structures may impact the compiler’s abilities to apply further optimisations, such as vectorisation for other architectures.  The building block notion is useful for simplicity reasons in further sections. In general, the compiler might have been able to apply these optimisations or similar ones in common architectures, but Loki’s compiler is not very efficient for these applications at the moment of writing and they have to be applied manually. One more reason for keeping the simplicity of the code snippet is that I evaluate parallel performance using OpenMP and irregular loop structures are not exploitable for parallelisation. 

\section{Access Pattern Manipulation}

One very interesting optimisation to explore is to change the access pattern so the working set could be decreased. By minimising the working set, the cache performance can be improved at a considerable amount, which can impact the overall performance. This is because by accessing the data in different fashions the spatial and temporal localities change. 

Other factors that can impact performance when changing the access pattern are the block localities, as an access to a memory reference requests the fetch of an entire block, which has a size of 32 bytes in the Loki architecture, and prefetcher’s performance for other architectures that support it. Loki supports manual prefetching through fetch commands placed inside the application’s text segment. While the block access can virtually act like a form of prefetching, it could also be a reason for fetching irrelevant data or with data with low-temporal locality. Some processors support the critical word first optimisation which could make the results more complex as there are additional block buffers \cite{24}.

The way I change the access pattern is by selecting different permutations of the loop blocks. In Figure 3.2 we saw this block notation that kept the building blocks simple. I selected this loop structure also because it is very easy to apply permutations. The code will produce correct results for every permutation out of the 720 ones for the 6 loop blocks, as long as the closing sections are in the correct order, which is the inverse of the loop permutation. 

Currently, due to the high number of combinations, there are no mechanisms or research work for exhaustive analysis or for deciding the loop order automatically, to my knowledge. Most of the loop reordering research topics are focused in maintaining the data dependencies of arbitrary nested loops automatically and how it can be scaled on parallel machines. The work “automatic loop interchange” \cite{30} summarises the concepts of loop reordering from older works and provides an algorithm for safe loop transformations for vectorising compilers.   

In Figure 3.3, we can see a visualisation of the address and block reuse patterns for a specific network configuration  under the best and worst loop permutation. The figures are for the first 100 million instructions of the execution of this configuration. The simulation framework for this, as well as how the worst and best loop order has been found is explained in Chapter 4. The results here are only for demonstration purposes. On the left side we see the reuse patterns for the best loop permutation, while on the right side we see the equivalent patters for the worst performing loop order. The top graphs represent the address reuse patterns which is also architecture independent in regard to the block size. The lower graphs represent the equivalent graphs for the block references instead of the address references. Some important observations are that the block reuse patterns are more representative for measuring performance and that the best loop order has a smaller working set in general. The worst loop permutation case has a very low address reuse in this time frame, but there is a considerable block reuse, not as much as the best case though. 

\begin{figure}[h!]
\centering
\includegraphics[width=1.02\textwidth]{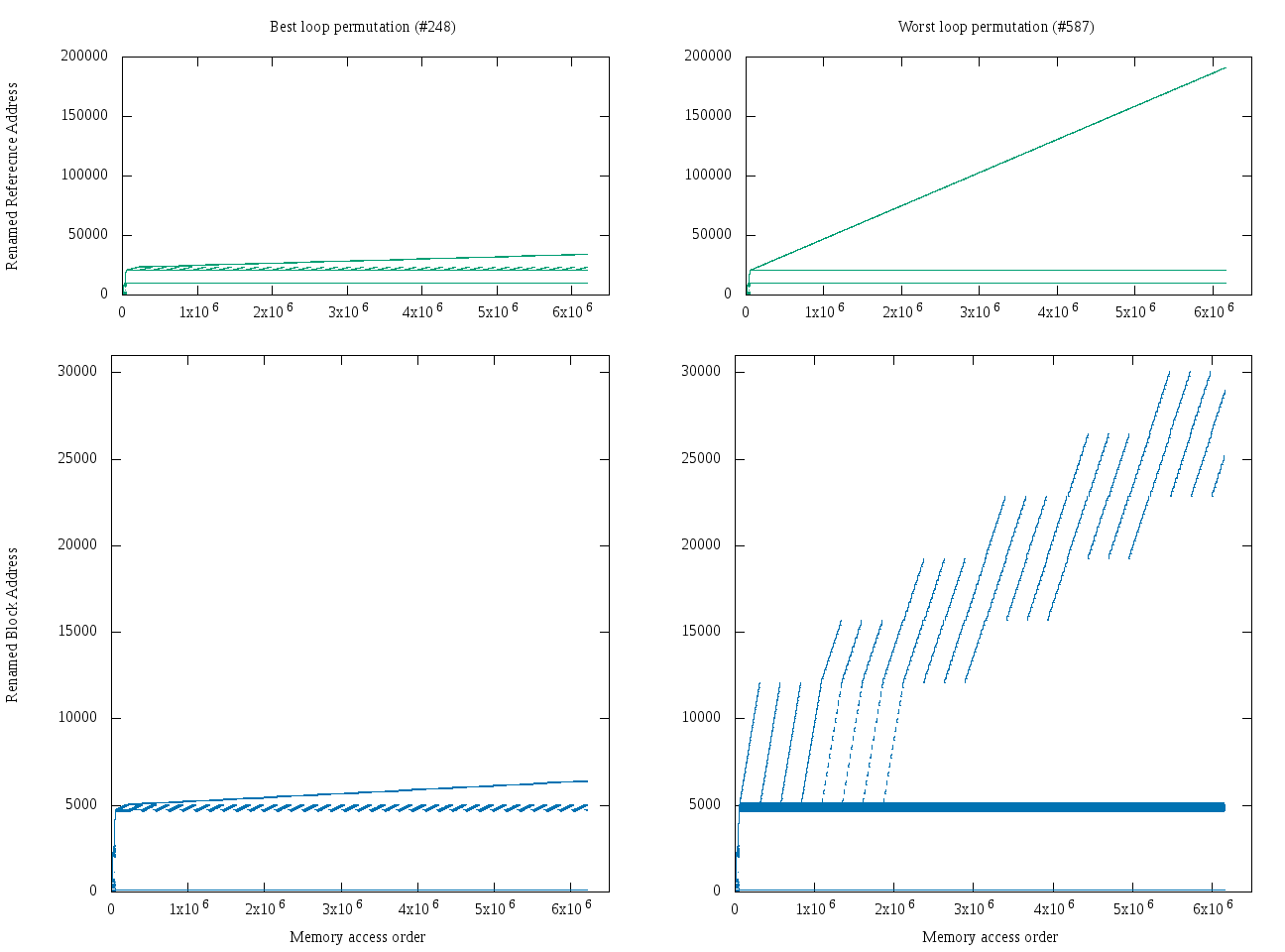}
\caption{Visualisation of the address and block reuse patterns. Initially two binaries were produced, one with the best loop order and one with the worst for the specific layer parameters. The execution of the first 100M instructions was instrumented to obtain the memory accesses references. Because the virtual address space is too sparse for visualisation, the addresses were renamed according to their time of first appearance. In that way we produced the two upper graphs that show the address reuse in a compact format. In the two lower graphs are the equivalent but after removing the word offset from each memory reference. The word offset size is equal to 5 to mach Loki’s offset.}
\end{figure}

Visually, we observe that the working set for the best case is around 500 blocks, which is approximately 16kB. Therefore an L1 size of 16 kB could be enough for this case. For the worst case, the working set seems to be around 5000 blocks, which is around 160 kB. That means that it would probably depend on the performance of a higher-level cache, which would have more access time than L1. It is important to note that there is not any formal definition for the working set as it is used to describe the data reuse in a short period of time.

\newpage
\section{Partial sums}

Due to the nature of the convolution algorithm, as with matrix multiplication, each entry inside the output layer is accessed multiple times for adding up a finite series of numbers. This requires a lot of memory writes which are much more expensive than read operations or writes to registers. We could eliminate the output array accesses by making this addition externally or at a local memory, such as inside a register or a scratchpad memory for Loki. Each sum can be computed partially at different stages and the final result could be computed later by adding partial sums or a single summation variable for the single-thread case. 

In Figure 3.4 I present the application of the partial sums optimisation for our base loop permutation. In the shaded areas we can see the parts that the summation command from the previous code has partitioned into. The dashed arrows show the data dependencies of the index of the out array. This essentially means that the code inside the last dependency loop writes to the same place in memory. For example, as we can observe from the given loop permutation of the figure, the code inside the 4\textsuperscript{th} loop writes on the same variable but produces the same result as the unoptimised code. 

\begin{figure}[h]
\centering
 \includegraphics[width=1\textwidth]{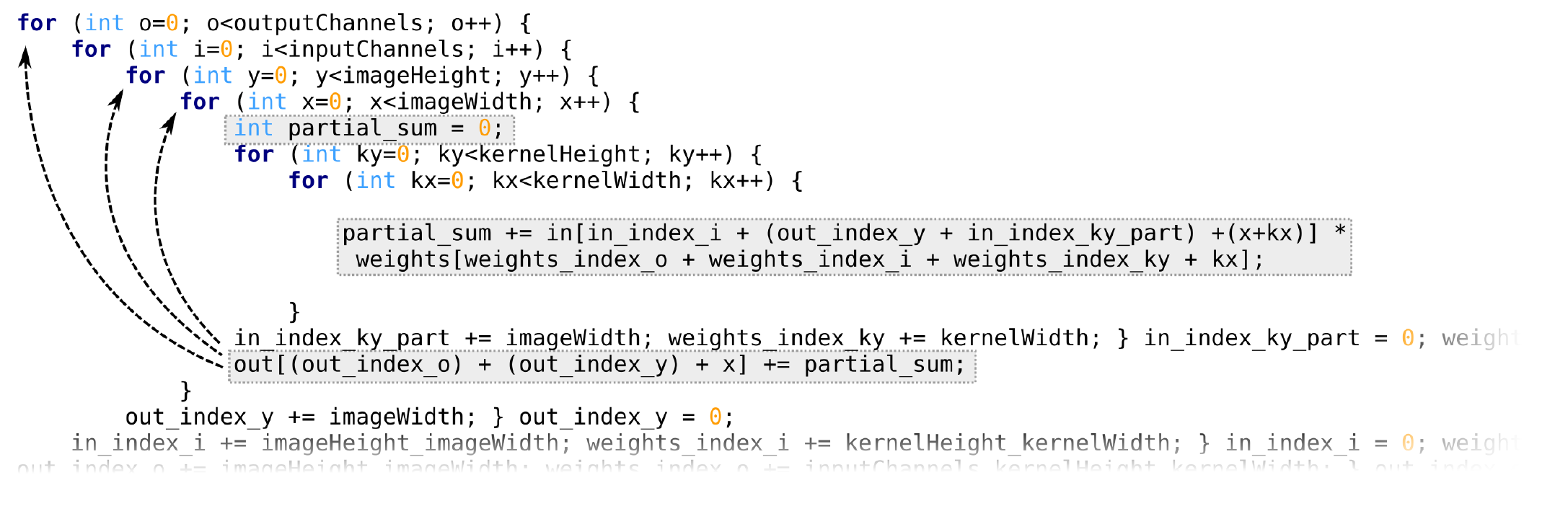}
\caption{Partial sums optimisation. The dashed arrows show the dependencies of the index of the out array.}
\end{figure}

It is interesting to note that this optimisation does not improve the reuse because the reuse is already high for the references to the blocks of the out array. The benefits are coming from the change of type of memory that the partial sum is on and the decrease of memory write operations.

\section{Parallel implementations}

As a first step for my evaluation of different optimisations and configurations I use OpenMP to parallelise the above code using the outermost loop. OpenMP is a straightforward parallelisation framework that enables effortless and efficient parallelisation of common programming models, such as the parallel for. At a later stage I validate the findings using an equivalent code for Loki, which is written manually. 

The main overhead of the parallel implementations usually comes from the mechanisms with which the thread safety is retained on the data structures that the threads use for writing shared information. In our case the data structure we need to protect will be the out array. Because the coherency measures are expensive, the partial sums optimisation has even more importance in the parallel case. The partial sums eliminate the times each thread is writing to the shared array and therefore fewer updates need to be monitored. 

OpenMP provides different thread safety measures, such as critical sections, atomic operations and locks. Experimentally, I have found that atomic operations are ideal for our case and the the critical sections are unnecessarily expensive for updating a single location in memory. There is a lot of research work for improving and validating the safety measures in both hardware and programming models \cite{31,32}. At this stage I wanted to keep the code simple and portable in order to make quick observations from the higher-level cache simulator.

In Figure 3, we observe the parallel version of a different permutation. The shaded regions are the changes from the serial version. First we insert the pragma to parallelise the outermost loop with the static scheduling option to minimize any run-time overheads for thread scheduling. Dynamic scheduling is probably unnecessary because we already know that each iteration carry an equally sized workload. The next difference is that we insert the atomic operation pragma above the out array update line. In order for it to produce correct results we also need to make all the iteration values private as each thread will need to have their own states of the iterations. For this reason I declare these private values at the first line inside the outermost loop. Last, the iteration values optimisation cannot be applied for the outermost loop because I declare them inside the iteration and therefore I insert the respective multiplications to the outermost loop. Of course this could be prevented by declaring those variables outside the loop structures and declaring them as private variables using inside the OpenMP parallel for pragma, but I find this more elegant for demonstration. 

\begin{figure}[h]
\centering
 \includegraphics[width=1.1\textwidth]{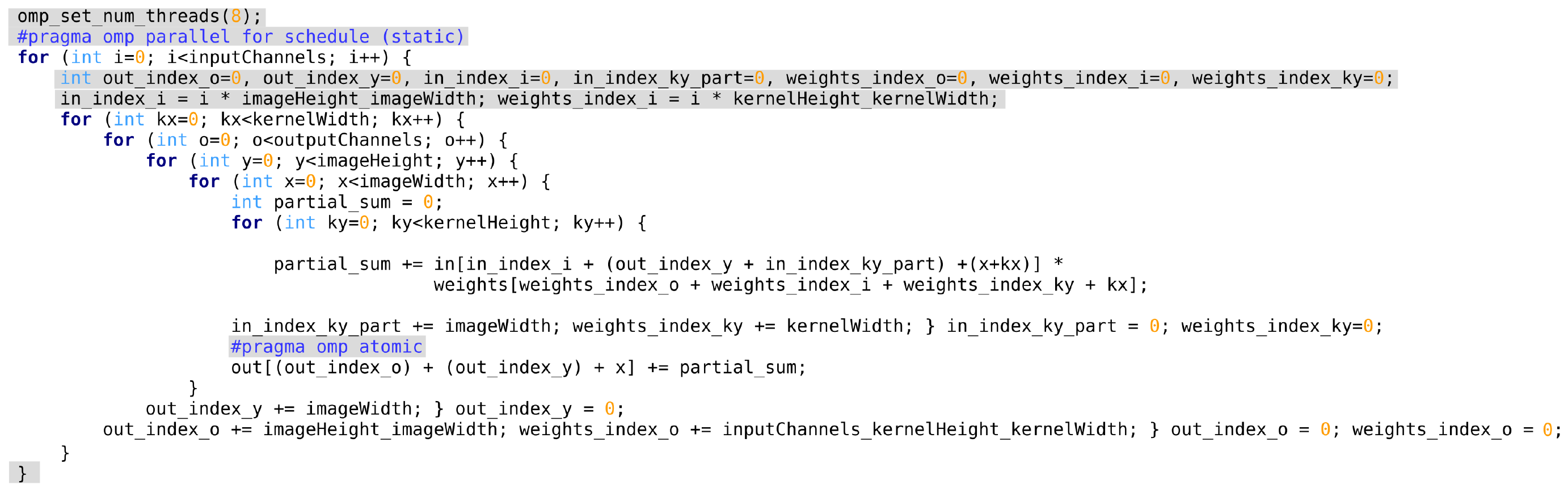}
\caption{Parallel implementation. The shaded areas are the changes from the serial version.}
\end{figure}

Another optimisation that is useful for completely removing the need and overheads of thread safety is to partition the iterations in a way that at any given time no two threads will write to the same memory location. As we have noticed, the index of the out array in the code is dependent to the values of o, x and y. This essentially means that we can safely remove the atomic operation pragma in every case we parallelise any of the loops that iterate using one of those three parameters. For example in the permutation of Figure 3.5 we cannot remove pragma omp atomic because the threads write on all location of the out array, while if we used the parallel implementation of permutation in Figure Y we could safely omit the atomic operation pragma.

One of the ideologies of scalable architectures such as Loki is to keep components simplistic. Therefore a design decision was not to include any hardware coherency mechanism in the memory hierarchy. Instead, there are cache flush and invalidate commands that have to be placed manually by the programmer or compiler. As a result, there is an extra effort to enforce thread safety on the Loki implementations. Of course, one easier solution is to prefer the permutations that eliminate the need of the thread safety measure by writing in to different segments of the output layer. In addition, in Loki you can map all 8 cores of a tile to a unified L1, consisting of all the tile’s memory banks, which simplifies the thread safety measures across the cores of the tile. 

A design space I will not explore in this project is the effect of parallelising an inner loop instead of the outermost loop. The benefit of the parallelisation of inner loops could be the better and more fair utilisation of the assigned the threads, if the outer loop consists out of a small number of iterations. Another result would be that the iterations of the outer loops would act as barriers and therefore more synchronisation would be enforced across the threads. More synchronisation could also eliminate the working set, because the threads would iterate over data with more temporal locality. Reasons for decrease in performance would be the extra wait time of the threads that have been given more compute time and finished earlier, as well as any additional OpenMP-generated overheads from the presence of more barriers.

\section{Convolution implementation generator}

In order to evaluate a big number of hardware configurations, input parameters and loop permutations I have created a python script that produces a C program according to the given parameters. The parameters are the input parameters, which are the convolutional neural network layer characteristics, and the permutation index, which ranges from 0 to 719 for the 6! possible permutations. The input parameters are the number of input and output channels, the width and height of the image, and the width and height of filter. This permutation index is based on the python itertools library, which produces an iterator that produces all permutations in a lexicographic order.

The output of the script is a single .c file which can be compiled by gcc and other compilers. The resulting code includes all the above optimisations, including partial sums and multiplication elimination. The ending sections of each loop is placed in the reverse order of the loop permutation automatically for the algorithm to work correctly. There is also a hardcoded variable that expresses the number of threads. When the threads variable gets a value higher than 1, it introduces the OpenMP-related code and also makes all the required changes demonstrated in Figure 3.5. It also removes the atomic operation pragma when thread safety is implied by the loop permutation.

The generated programs iterate over arrays filled with zeros to eliminate data loading times. I also use malloc instead of calloc to request the memory pages only on-demand because I would like to isolate and explore the memory access patterns of each loop permutation. When compiling with gcc, the optimisations should be turned off with the flag -O0 because the gcc optimisation heuristics remove almost all operations of the code as we do not initialize the arrays or use the output layer. In combination with the manual non-architecture-specific optimisations, I examine realistic access patterns isolated from the array initialisation phase which would be already implied in a CNN convolution layer. I have also created a version that inserts random data on all implementations and loop permutations and ensures that the result is the same in all combinations for validation purposes.

In Figure 3.6, I present a set of preliminary results using 1 thread on real hardware, a Haswell machine. The code is compiled using gcc without any optimisation and each configuration is run 50 times. The plotted values are the respective median, to eliminate outliers as this is not an isolated environment. The best loop order is found by running all 720 permutations and selecting the best performing one among the medians of 20 runs per permutation. The y axis shows the kind of the applied optimisations, where constants and iterations are the multiplication elimination measures described in 3.1. The only optimisation of these that needs tuning is the best loop order selection and here the result is not very promising as it offers around 40x speedup. One reason for this is that the initial loop was among the top as the worst loop order had around 3 times slowdown and maybe the selected input parameters produced less demanding workload. In the loop reordering chapter I also will explore the impact of the different configurations on the top in a processor with lower LLC cache size and multiple threads.

\begin{figure}[h]
\centering
 \includegraphics[width=1\textwidth]{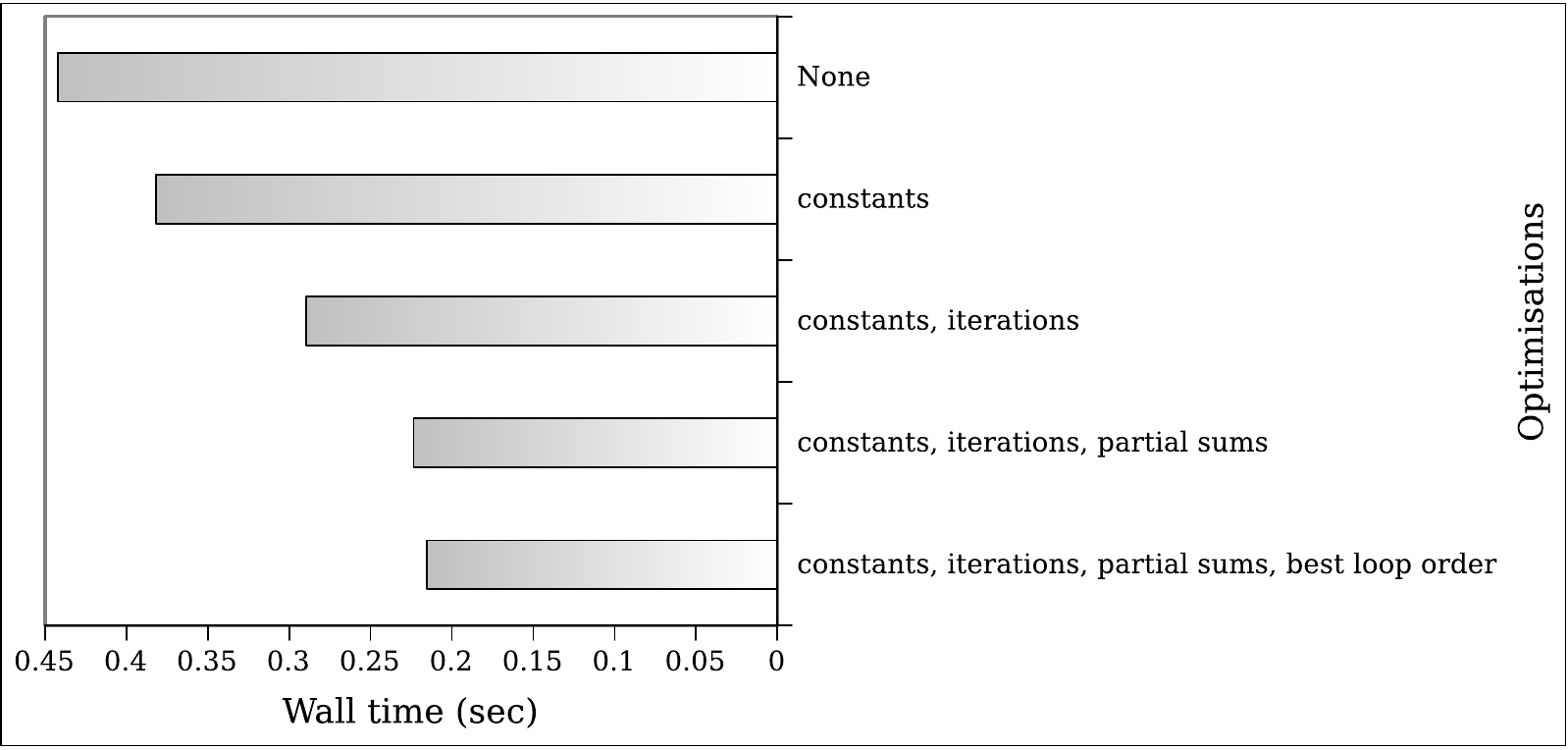}
\caption{Preliminary experiment to show the execution time after applying a variety of optimisations. Single-thread execution on an x86 server.}
\end{figure}

\section{Sparsity-sensitive algorithms}

One algorithm optimisation that is interesting to explore is related to the sparsity measures. The general idea is that when the multiplier or the multiplicand is equal to zero, then the addition of their product to the output layer is unnecessary. In the main summation operation we note that the two values that could equal to zero are the weight and input image data. Therefore, we can implement the optimisation by adding checks to skip all iterations in which one of the multiplicand or the multiplier is equal to zero. This usually saves computation time, but the performance is highly dependent to the actual sparsity of the input layer and the filter layer.

The convolution implementation generator gives the same performance across all input layers and filters. This is not the case for an activation-aware or weight-aware algorithm. For example, if one image has only 10\% activations the sparsity-sensitive algorithm may be more efficient than the dense algorithm. If the image has many zeros in the input layer, then the dense algorithm will certainly perform better as it does not have the checks which would otherwise still not save computation. 

Generally, when evaluating this implementation variation it would be very important to make a case study of the common ranges of percentages of sparsity in the image and the weights. It would also be useful to consider the distribution of the dense areas in such cases. If some regions are very dense and the loop permutation assigns an 100\% dense region to a single core, then this could become a bottleneck. Therefore we must be very careful when making generalizations about the sparsity measures.

The focus of my convolution implementation generator is mainly to explore access patterns in a cache simulator without specific input values and therefore I do not implement the sparsity measures. The Loki team has implementations that add sparsity measures for the weights and activations and I will evaluate the results directly on the lokisim framework in a later section. One other reason for exploring them directly on lokisim is because they might further improve performance if there is a bandwidth bottleneck across the Loki tiles and the results of the cache simulator would not be as representative. 

Something interesting to note is that on a real-world OpenMP implementation of the parallel case with the sparsity sensitivity optimisation it might prove essential to use dynamic scheduling instead of static thread allocation as different iteration chunks can have different amounts of workload, according to the distribution of the zeros in the image and filters.

\section{Loki-specific and other optimisations}

There are many other optimisation techniques that could be applied, but they can be architecture specific. One example is loop unrolling. The automatic loop unrolling of the compiler of the innermost loops could change the innermost loop structure in order to eliminate the jump assembly commands and the related data dependencies and further increase the performance. 

In Loki implementations this optimisation is currently done manually, as the compiler is immature and programmer’s knowledge of the architecture is essential. Each core in the Loki processor has two kinds of instruction caches, instead of the regular private instruction-only L1 cache. There is one 64-word instruction packet queue (L0) that stores instruction packets (IPKs) containing the basic blocks of the program and also one higher-priority 16-word instruction buffer, that aims to save power and increase performance for repetitive tasks \cite{2}. During manual optimisation, the programmer acknowledges the size of this buffer and can unroll the loops up to a specific number of commands and minimise the instruction misses.

The Loki architecture provides the possibility of adapting to different memory hierarchies and specializing the resource allocation to the application's needs. The basic form of this functionality is to disable a specified number of tiles and use all their memory banks as a shared L2 cache. Therefore there is also an additional parameter for optimisation which will be explored in Chapter 6. Ideally the application will use a substantial amount of cores for high throughput and provide the amount of cache that will both eliminate the number of LLC misses with a block request rate that will not flood the interconnection network and cause stalls. Of course, there is also the question whether there will be an optimal configuration provided the architecture constraints and the wide range of possible inputs for the convolutional neural network application.

One other feature of the Loki architecture is the ability to partition the L1 cache into custom-sized memory bank groups for private use by selected cores or for different data structures. Since the L1 cache is direct-mapped, the partitioning of the 8 available memory banks can help reduce conflict misses. Another benefit of the L1 cache partitioning is that it can provide quality of service when some data is predicted to have high reuse or they have other properties that would otherwise degrade performance \cite{22}. Examples of those properties are thrashing and scanning cache access behaviours, where the working set is too big to fit in cache or the data is accessed in a stream fashion and has no reuse \cite{23}. In the latter case, Loki is able to programmatically bypass L1 or perform direct memory accesses. 

One other optimisation that is applicable on the main convolution code is loop vectorisation, if the architecture supports it. On conventional multicores the architecture has dedicated assembly commands to support Single Instruction Multiple Data (SIMD) operations. The compiler checks which set of instructions are available to the host machine’s processor and can perform automatic vectorisation of loop structures for a more efficient utilisation of the available processor resources. Due to the packed-based architecture, Loki allows the remote assignment of specific tasks to specific cores during run-time. This is useful for implementing Data-Level Parallelism. Some cores can be set as helper cores and distribute similar workload to the rest of the cores. The helper coreas can also be used for load balancing. One design question would be to find the best performing ratio of helper and worker cores. Again, all these options contribute to the very wide design space of exploration of software optimisations. The reason that they can be considered software optimisations for Loki is because these options are exposed to the instruction set architecture.

\chapter{Loop Order Analysis}

In this chapter I present my analysis of the nested loop permutations in the most time consuming nested loop section of convolution in a CNN. The idea is to alter the access pattern in order to minimise the working set used during the calculations and therefore eliminate the cache misses. By eliminating the number of both L1 and L2 misses the performance is shown to be improved by a significant amount, not only due to the respective L2 and main memory access latencies, but also due to the reduced congestion in Loki’s interconnection network, which is sometimes proven to be a bottleneck in performance.  

By permuting the main nested loop section we can find an optimal permutation that increases both the time and spatial localities in memory accesses. The question that will be answered at at later stage is whether there is a specific or a set of loop order permutations that perform near optimally on all input cases and cache configurations. One other important question to be answered is how critical is this decision and if is proven to be important, how an on-the-fly loop re-ordering mechanism could be implemented in software.

\section{Experimental setup}

In order to evaluate all the number of loop permutations (in our case 6! = 720) on many configurations efficiently, I have used my custom cache simulator. The x86 binary is produced by my convolution generator script. The arguments for a single binary are the permutation index and the input parameters (the number of input channels, output channels, the width of and the height of the image and the width and the width of the kernels). The architectural parameters were summarised in Table 2.1 at the methodology section of Chapter 2.

\section{Hamiltonian path index for permutation visualisation}

An innovative use of the Steinhaus-Johnson-Trotter algorithm (1963) \cite{21} is presented to represent the different permutations in an order that is based on spatial characteristics. The idea is to find an indexing function for all permutations using one parameter that carries some locality information. In this way we could distinguish regions of loop permutations that perform better than others and compare signatures more effectively, or even use it for dynamic loop reordering if we try to optimize a single parameter. For 6 elements the number of permutations is 6!=720 and the number of ways the indexing can be done with a single parameter is 6!! = 720!, which is an 1747-digit dumber. 

Examples of common indexing functions are the lexicographical order and the reverse lexicographic order. This can be a bad idea for visualisation. In the first case for example, the permutations (4, 5, 3, 2, 1, 0) and (5, 0, 1, 2, 3, 4) are consecutive but they look very dissimilar. However, the fact that the more left an element is the less rapidly changes among consecutive permutations gives it another locality property. This property could be useful if we knew for sure that one end of the permutation has more impact on the dependent variable. In our case, this might hold for L1 misses, as innermost loops may determine the “immediate” working set, but for L2 misses this might not be the case.

The design space when dealing with permutations can be represented as an undirected graph of which each node represents a permutation and each edge connects the permutations that differ only by a neighbour (adjacent elements) swap. A simpler graph for n=4 can be seen in Figure 4.1. For n=6, the number of nodes is 720 and the number of edges is 1800. By using the Steinhaus-Johnson-Trotter algorithm we can get a hamiltonian path (a path that visits each node exactly once) of this graph which can then be used as an alternative indexing function.

\begin{figure}[h]
\centering
 \includegraphics[width=0.4\textwidth]{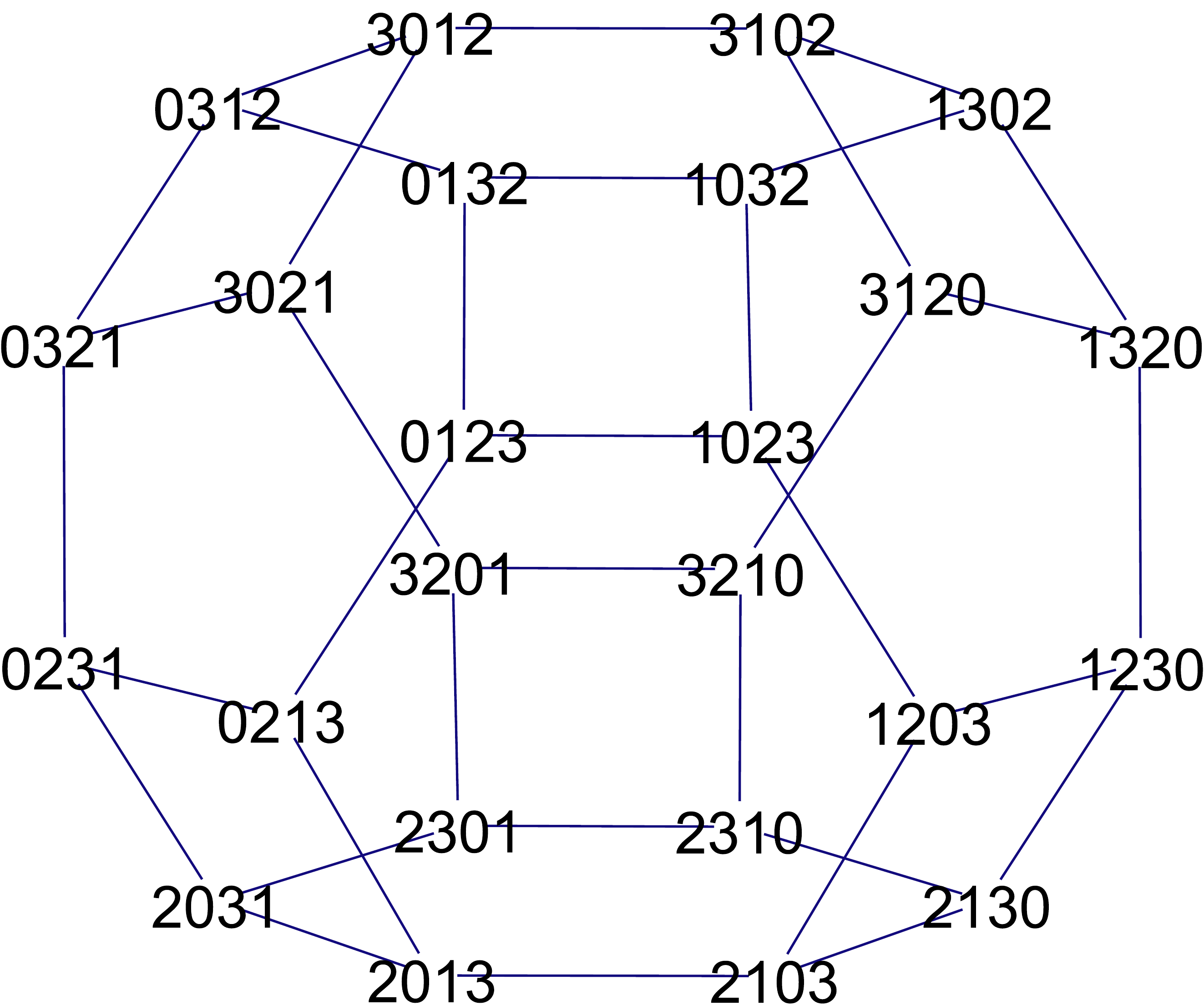}
\caption{A permutohedron where the nodes are all permutations of 4 elements and the edges connect the permutations that differ only by an adjacent elements swap.}
\end{figure}

In Figure 4.2 I compare the results of 720 simulations, one for each permutation for the same input parameters, in 3 different orders. The first two indexing functions are the lexicographic and reverse lexicographic orders, as implemented in the python programming language library called itertools that provides iterators on permutations of object lists. As we can see, the hamiltonian path index is a very good option for visualizing permutations in the 3 evaluation metric cases. Another interesting observation is that there is some periodicity at the hamiltonian path index graphs and that is because we lose some locality information that would have probably existed in the initial graph. 

If we compare the graphs individually, one of the lexicographic orders for the cycles graph produced a less noisy graph than the other indexing methods. However, this indexing might not be good as a general solution if we are not aware of any importance of the leftmost or rightmost elements of the permutations. In addition, the hamiltonian path index tends to produce the most distinguishable regions of hills and valleys of bad and well-performing regions respectively, which will later prove to be more suitable for graph comparison. The permutation graphs will be used as signatures for visual comparison of different input and configuration parameters. 

The last observation is that in the cycles case, the fact that the inverse permutation python index produced this kind of downward shape shows the importance of the innermost loops (rightmost objects in a permutation) for determining the working set. By using the reverse lexicographic order, the x-axis is partitioned into 6 equally-sized segments which contain 120 loop permutations which share the same innermost loop. 

\begin{figure}[h!]
\centering
 \includegraphics[width=1\textwidth]{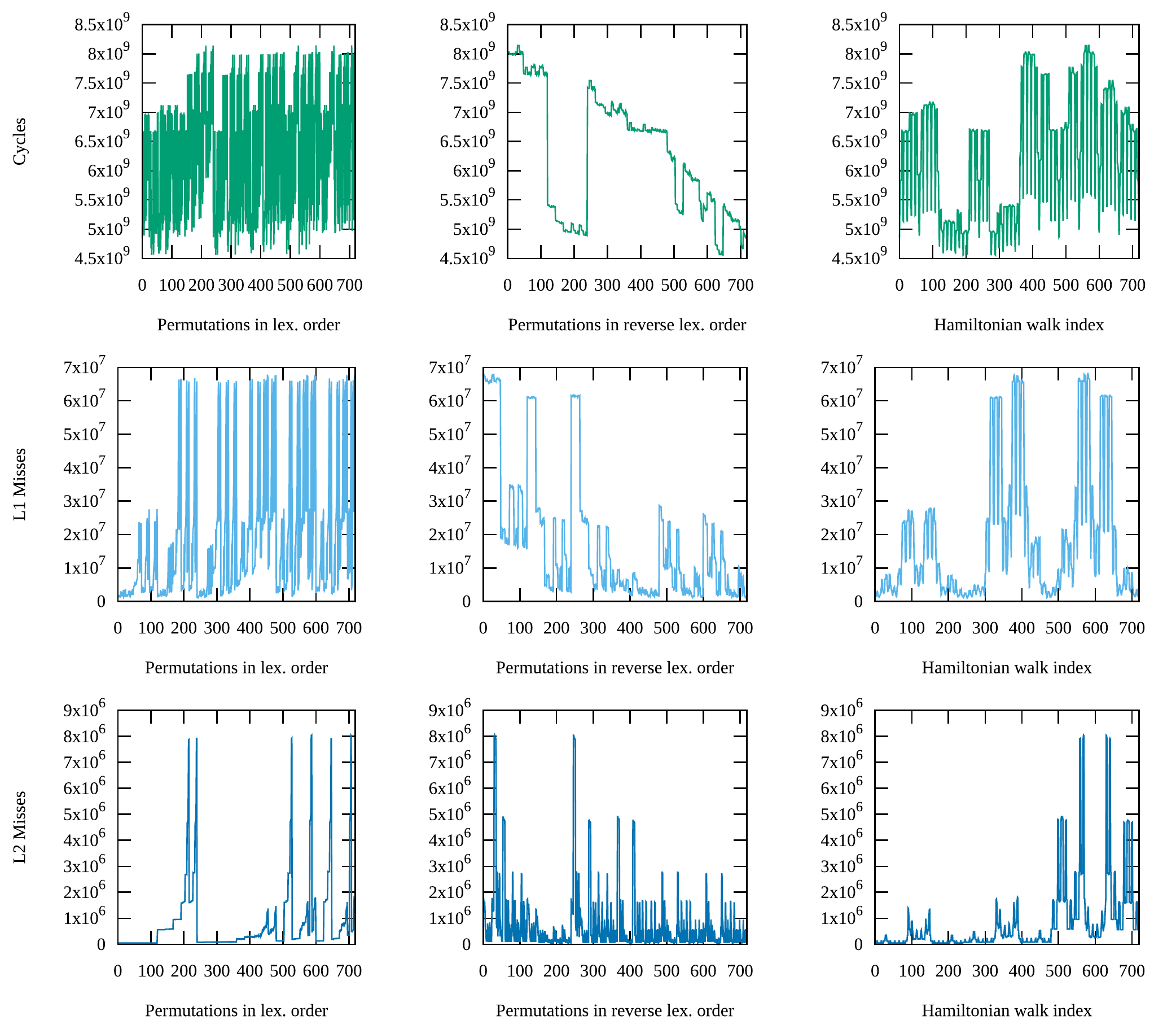}
\caption{Cycle, L1 misses and L2 misses results from 720 simulations, one for each loop permutation, using different indexing methods. Input parameters: 256, 32, 28, 28, 3, 3 for output channels, input channels, image width, image height, filter width and filter height respectively. Single-thread version. The layer parameters is the specification of the 10th layer of TinyDarknet \cite{25}.}
\end{figure}

The hamiltonian path indexing method could also be used in other applications where different permutations are explored visually or when searching for an optimal index and there are localities among neighbouring permutations.

\section{Comparing signatures for different inputs and configurations }

In this section I present the results of the simulations of different input parameters for all loop permutations. The input parameters are explained in section 2.2 and they all contribute the overall problem size. This section is divided in two subsections that are the analyses of different ranges of inputs. In the first, I explore the performance aspects of 7 convolution layers of SqueezeNet \cite{14} when changing loop permutations. In the second case I present a wider range of synthetic layer parameters for the purpose of generalization. One of the aims of this section is to prove or disprove the need for dynamic loop interchange. This result will only be indicative because the simulator for this section is the cache simulator, which has a high level of abstraction and approximations.

\subsection{Layers from SqueezeNet and Tiny Darknet }

Squeezenet \cite{14} is a convolutional neural network topology designed with portability in mind. The architecture includes 10 main convolution layers in total and they constitute the majority of all steps. I have explicitly selected 7 layers of SqueezeNet, as well as one layer from Tiny Darknet \cite{25} to create a small design space with real world combinations of input parameters a variety of ranges. Table 4.1 summarizes the selected input parameters.

\begin{table}[h]
\centering
 \begin{tabular}{||c||P{1.8cm}|P{1.7cm}|P{1cm}|P{1cm}|P{1cm}|P{1cm}|P{1cm}||} 
 \hline
Layer&
Number of Output Channels&
Number of 	Input Channels&
Image	Width&
Image	Height&
Kernel	Width&
Kernel	Height&
Source
\\ [0.5ex] 
 \hline\hline
initial-conf&256&32&28&28&3&3&\cite{25}\\\hline
fire3-conv3x3-2&64&16&55&55&3&3&\cite{14}\\\hline
fire4-conv1x1-1&32&128&55&55&1&1&\cite{14}\\\hline
fire4-conv1x1-2&128&32&55&55&1&1&\cite{14}\\\hline
fire7-conv1x1-1&48&384&27&27&1&1&\cite{14}\\\hline
fire9-conv1x1-1&64&512&13&13&1&1&\cite{14}\\\hline
fire9-conv3x3-2&256&64&13&13&3&3&\cite{14}\\\hline
conv-final&1000&512&13&13&1&1&\cite{14}\\ [1ex] 
 \hline
\end{tabular}
\caption{The parameter values for the selected convolution layers.}
\end{table}

By observing all the created signatures of Figure 4.3, we can conclude that there certainly are some regions of good permutations across all layers. The two valleys around the permutation indexes 200 and 300 seem to perform well in all cases in this 1-thread experiment. Another observation is that among the lower performing permutations there is greater variation when comparing the signatures of different layers. The reason that the signatures of the second and third column seem less noisy is because the filter size is 1x1 and the displacement of the kernel width and kernel height do not have a significant impact on performance, since they only do one iteration. 

On average there seems to be around 2 times speedup from the worst loop permutation to best, although we are more interested in permutation comparison in this step because the cycles notation comes from the cache simulator, which is mainly a functional simulator.

\begin{figure}[h]
\centering
 \includegraphics[trim=2cm 0 0 0, width=1.1\textwidth]{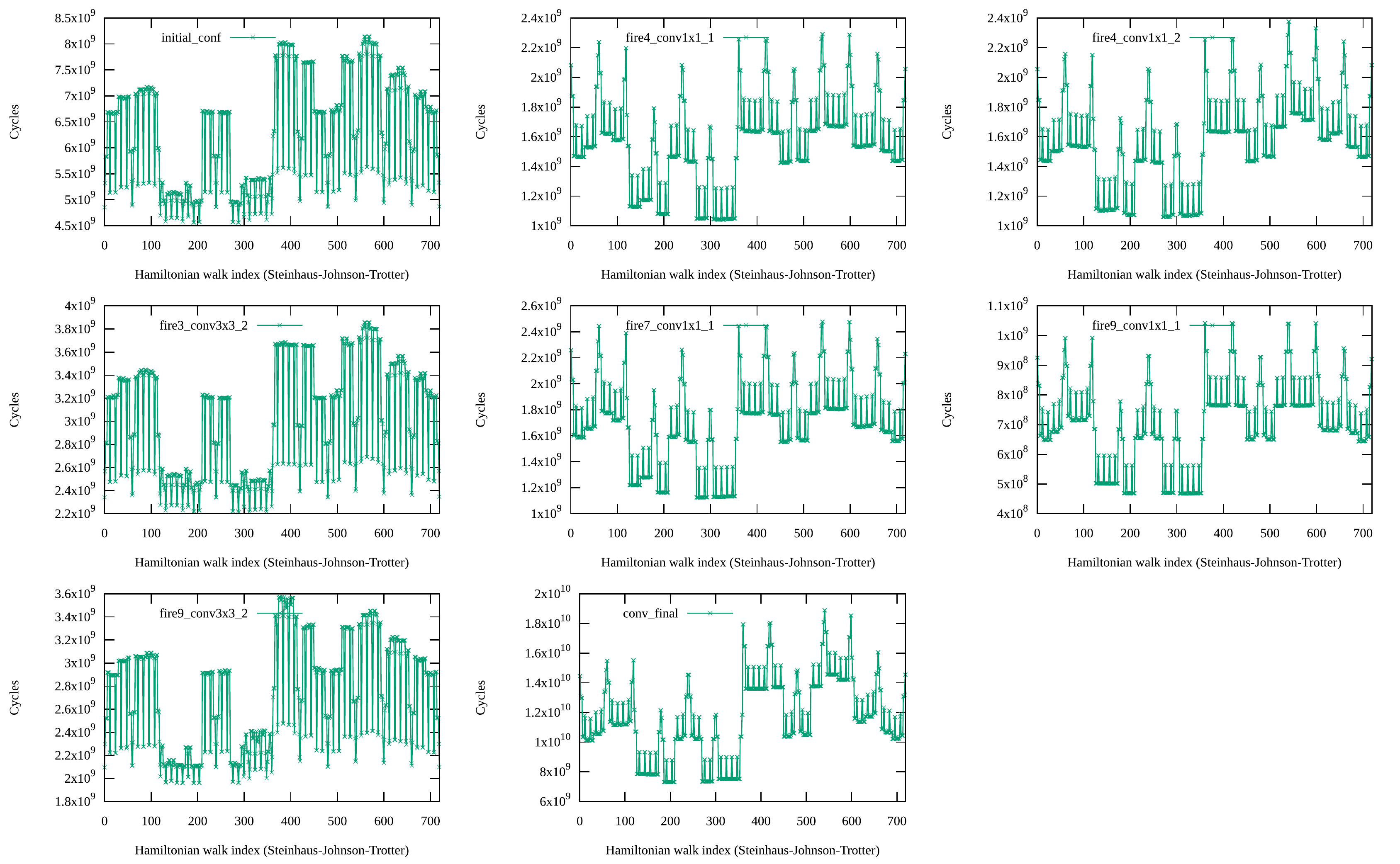}
\caption{Cycles for each layer of the small design space for each permutation. The permutations sorting is based on the Steinhaus-Johnson-Trotter algorithm \cite{21} to identify patterns visually.}
\end{figure}

In Figure 4.4, we can see the multi-threaded results. One observation is that there are good permutations more frequently as the number of threads increases. There seem to be cases of superlinear speedup. The reason for that might be that the threads help each other by prefetching useful data. As the workload is very similar across the threads there should be high probability of iterating over similar data and blocks in general. In the 1x1-sized kernel layers, we can observe all the cases that the kernel height or kernel width is the outermost loop. This is because OpenMP parallel for does not exploit any parallelism from 1 iteration loops. Small kernels could be considered a bad option for the multi-threaded design space because of the limited parallelisation they offer. However, it is already desirable to consider the kernel loops as outermost loops to be in bad permutations due to their limited exploitable parallelisability in the general use case.

\begin{figure}[h]
\centering
 \includegraphics[trim=2.4cm 0 0 0, width=1.1\textwidth]{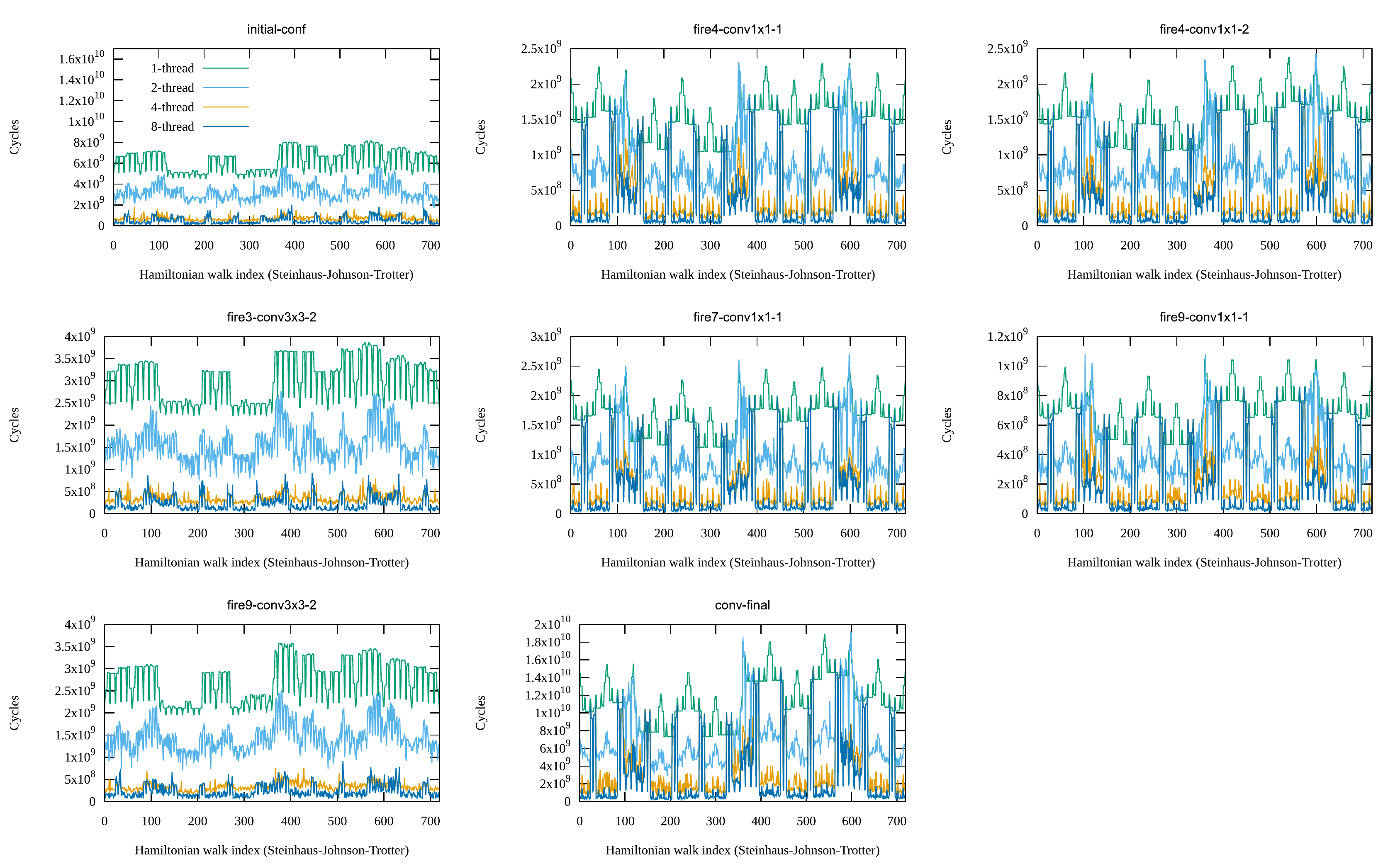}
\caption{Cycles for each layer in 1,2,4 and 8-thread modes for each permutation. The permutations sorting is based on the Steinhaus-Johnson-Trotter algorithm.}
\end{figure}

As a validation I have also measured the L2 misses for each of the different layer experiments (Figure 4.5). For the multi-threaded versions the L2 misses scale sublinearly while increasing the number of threads for the results obtained from the 1-thread versions. This is because there are many shared blocks. There are some exceptions though, in which the multi-threaded cases have many more misses, because of conflict and capacity misses (layers fire4-1 and fire7 at the region between around 100 and 150). This is because in a subset of permutations, different threads can write or read in different regions of the arrays. This selection includes the cases where we would consider them advantageous because we could omit the thread safety measures, but as we can see from here they also also have more demands from L2 cache. Also, in some cases such as the biggest trough in the fire7 layer, more threads can produce less misses. However, the L2 graphs alone does not give a clear picture of the L1 behaviour, which is more prone to noise from thread scheduling effects.

\begin{figure}[h]
\centering
 \includegraphics[trim=2cm 0 0 0, width=1.2\textwidth]{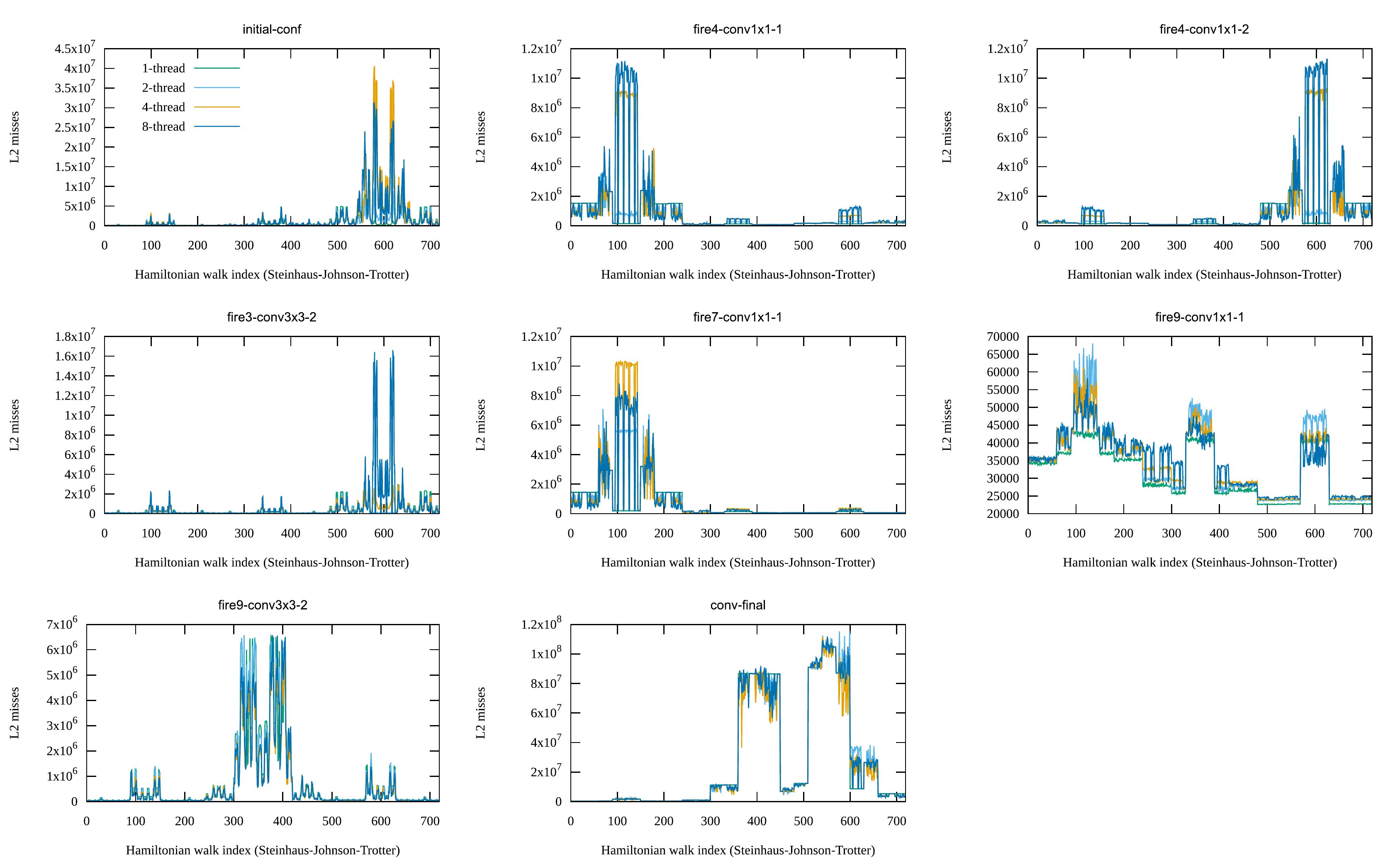}
\caption{L2 misses for each layer in 1,2,4 and 8-thread modes for each permutation.}
\end{figure}

\subsection{Synthetic layers}

In this section I present some results based on a synthetic design space to attempt generalisation for good permutations overall. In Table 4.2, I describe my design choice decisions. There are 216 combinations in total. Instead of variable grouping, it is more common to use other techniques for fast multivariate design exploration, such as the latin hypercube sampling (LHS), but here we already know that square images and kernels are a common case.

\begin{table}[h]
\centering
 \begin{tabular}{||P{3.5cm}||c|c|c|c||} 
 \hline
Variable Name&Lower Bound&Upper Bound&Increment&Total\\ [0.5ex] 
 \hline\hline
 Output Channels	and	Input Channels (equal)&10&210&40&6\\\hline
Image Width	Image Height (equal) &10&210&40&6\\\hline
Kernel Width	Kernel Height (equal)&1&11&2&6\\\hline
Total&-&-&-&216\\ [1ex] 
 \hline
\end{tabular}
\caption{The parameter ranges for the selected convolution layers.}
\end{table}

In order to save computation time I have limited the execution to 500 Million instructions. Usually, the representative region of execution is selected more carefully \cite{26} when limiting simulations. This is because the programs usually consist of many phases. In our case we have isolated only one phase of computation, convolution, which is a highly repetitive procedure. The methods of selecting the representative regions are ether by using automated routines \cite{26} or by visualising signatures of phases, such as with the number of touched program counter values over each N number of instructions. From my access pattern visualisation of section 3.2 we can observe that the initialisation phase is a very small fraction of the first 100M instructions (this was for 500M), at least for memory accesses, and then a relatively repetitive pattern continues until the end. The graph of the full execution is not presented for practical reasons.  

In Figure 4.6 I present a random subset of the hamiltonian path index signatures obtained from the design space exploration. As we can notice there are two main families of signatures. The most common one has 3 big troughs and 5 small valleys, while the less common has bigger valleys and they are skewed to the left. The second type is more similar to SqueezeNet’s signatures.

\begin{figure}[h]
\centering
 \includegraphics[width=0.7\textwidth]{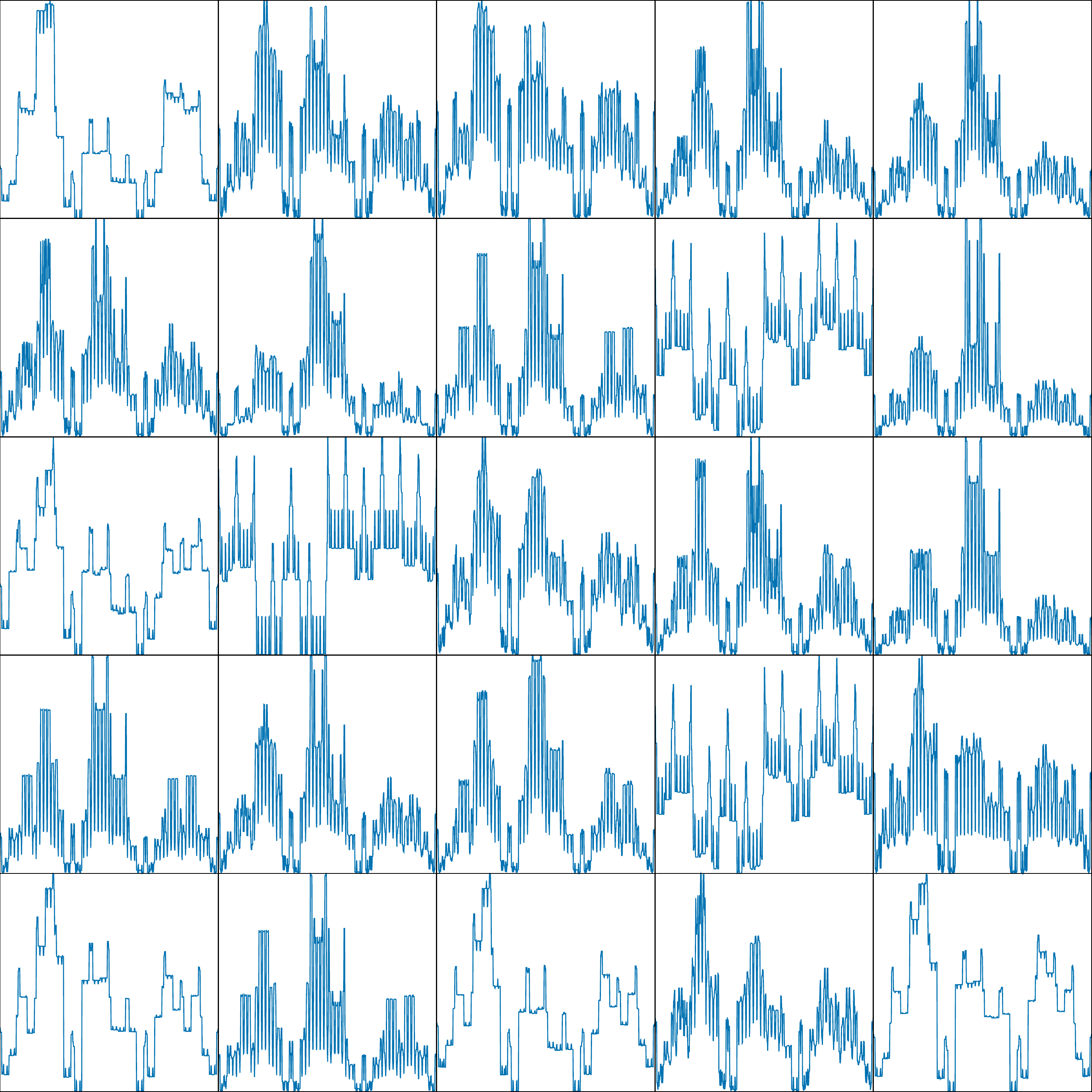}
\caption{Random subset of the cycle signatures for the bigger layer design space.}
\end{figure}

Finally, we would like to see if there is a common good performing permutation that could be used as a static choice in the convolution code for layers of which the parameter values are close to the respective ranges of the design space. In order to do this we find the best permutation per layer and then see how well each of the 720 permutations performed globally. As we can observe from the left graph of Figure 4.7, there is a permutation that performed 97\% optimally on average and 60\% at the worst case. There is also another one that performed 94\% optimally and 83\% at the worst case. These are two good candidates for evaluation on the Loki simulator.

\begin{figure}[h]
\centering
 \includegraphics[trim=1.8cm 0 0 0,width=1.03\textwidth]{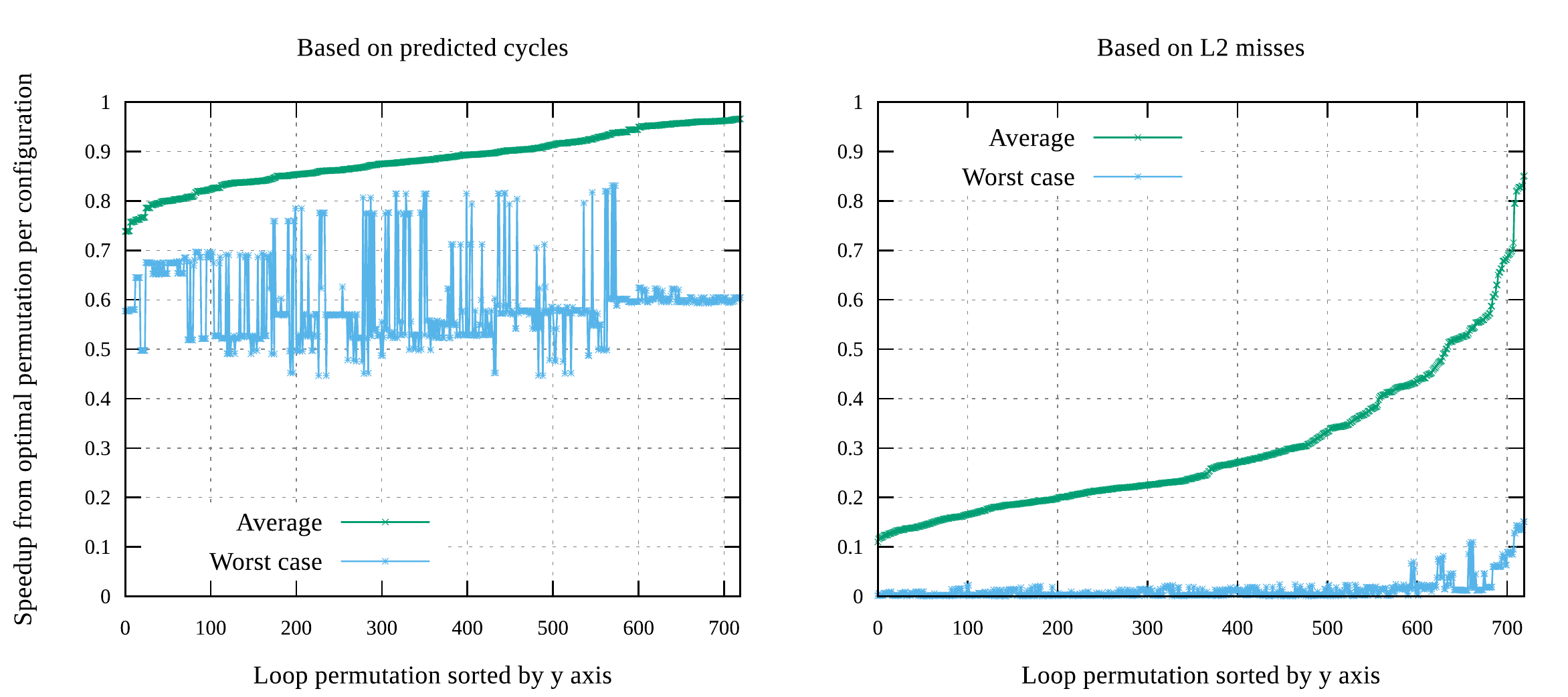}
\caption{Speedup each permutation achieves on every layer in comparison with the layer's optimal permutation, on average. On the Left, the average speedup is based on the cycles metric, while on the Right it is based on the L2 misses.}
\end{figure}

These results are based on cycles measurements. If we wanted to give emphasis on L2 misses because they can introduce link contention, we could also evaluate the equivalent candidates based on L2 misses. In the right graph of Figure 4.7 we can see that there is a permutation that has an average performance of 85\%, but the worst case is below 16\%. The top based on worst-case is very similar to the top average. If this candidate proves to perform better than the ones based on cycles, it may worth to consider dynamic loop reordering. 

The winner permutations for a single thread can be seen at Figure 4.8. There are some similarities between them. The outermost loop is common for the three loop permutations and is the image height. The three outermost loops are common for the left and right permutations, which represent the best based on cycles and best based on L2 misses. This could support the hypothesis that the outer loops are more important for determining the L2 cache performance, at least when the innermost loops do not produce a working set larger than the size of L2.

\begin{figure}[h]
\centering
 \includegraphics[trim=1.5cm 0 0 0,width=1.1\textwidth]{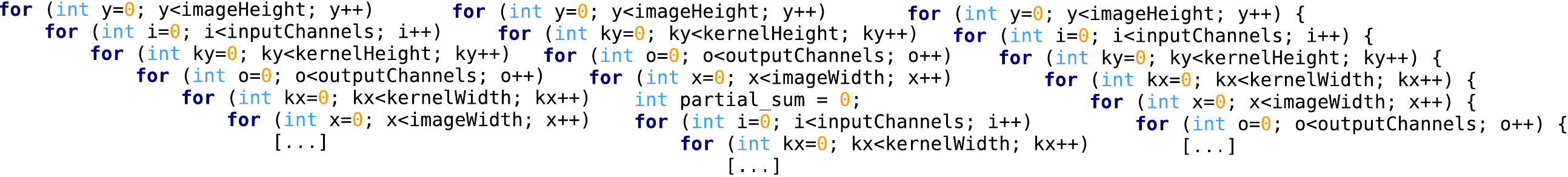}
\caption{The 3 candidates for best loop permutations for 1 thread. From left to right: Top average speedup (0.966004), Top worst-case speedup (0.831247, average 0.937533), Top speedup based on L2 misses (0.851068)}
\end{figure}

I have also explored an equivalent design space for the multi-thread case. The number of threads for this experiment is set to 8, which represents one tile. The ranges of the layers have been reduced due to time constraints, as well as the number of simulated instructions. The total number of layers are 36 and the instruction limit is set to 100 million instructions. In Table 4.3 we can see a summary of the layer design space.

\begin{table}[h]
\centering
 \begin{tabular}{||P{3.5cm}||c|c|c|c||} 
 \hline
Variable Name&Lower Bound&Upper Bound&Increment&Total\\ [0.5ex] 
 \hline\hline
 Output Channels	and	Input Channels (equal)&10&170&80&3\\\hline
Image Width	Image Height (equal) &10&170&80&3\\\hline
\multicolumn{1}{||P{3.5cm}||}{Kernel Width	Kernel Height (equal)}&\multicolumn{3}{c|}{1,3,9,11}&4\\\hline
Total&-&-&-&36\\ [1ex] 
 \hline
\end{tabular}
\caption{The parameter ranges for the selected convolution layers, multi-thread version.}
\end{table}

We also want to find a single top permutation that would perform well in the average case for multi-threaded experiments. Since Loki’s potential is based on the number of cores we want to exploit this opportunity for the multi-threaded convolution implementations. In Figure 4.9 we can see the results for each permutation. The main difference from the single-thread version is that now there is not a single near optimal loop permutation. When based on cycles, the top 1 is below 0.80 average speedup and the worst case is below 0.50. When based on L2 misses, the graph does not look very different from the 1-thread case, which means there might be a need for dynamic loop reordering for best performance.

We can also observe a big step at Figure 4.9 (left) from 0 to 239 and this is because exactly one third of the loop permutations have a kernel loop (kernel height or kernel width) in the outermost position. Because it is common to have small values for kernel width and kernel height, such as 1, the outermost loop in these cases is not parallelised sufficiently, with the result of impacting performance.

\begin{figure}[h]
\centering
 \includegraphics[trim=1.6cm 0 0 0,width=1.05\textwidth]{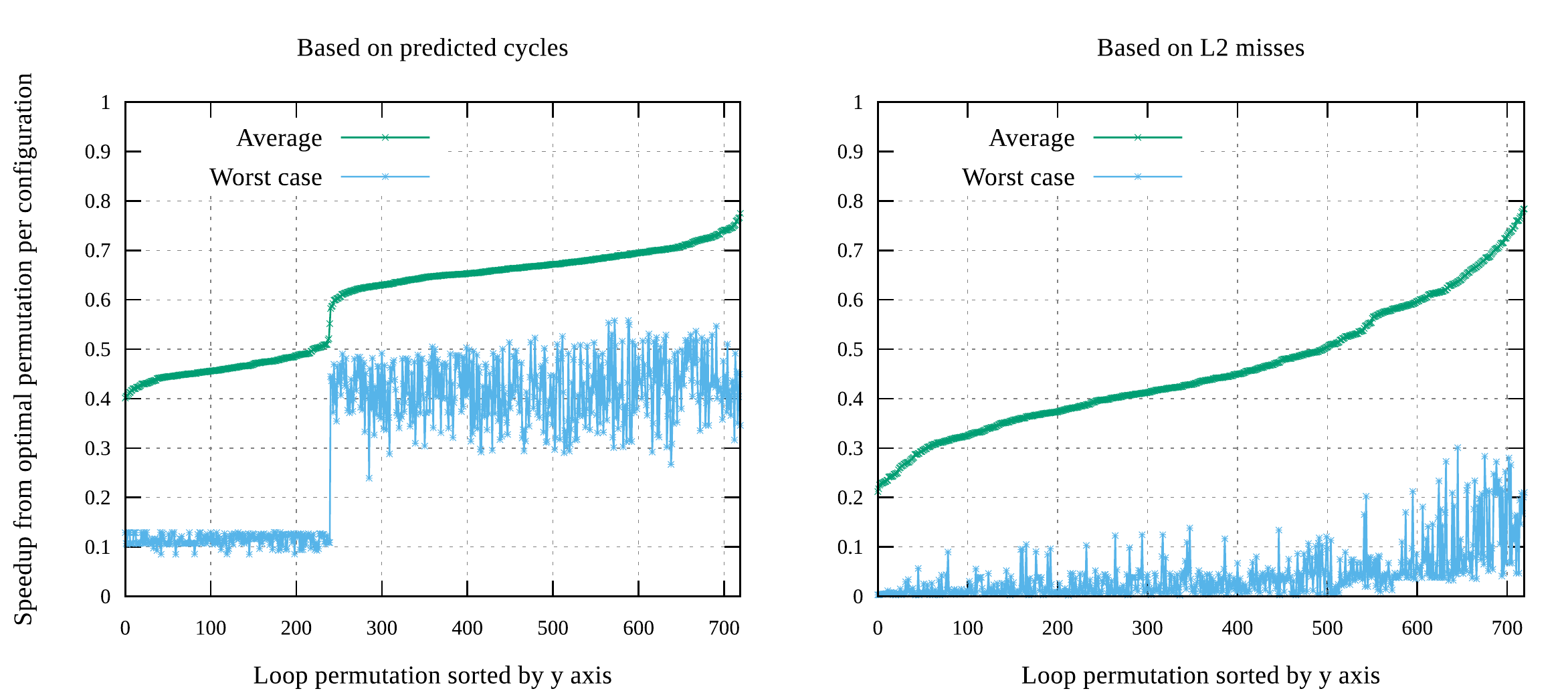}
\caption{(8-threads) Speedup each permutation achieves on every layer in comparison with the layer's optimal permutation, on average. On the Left, the average speedup is based on the cycles metric, while on the Right it is based on the L2 misses.}
\end{figure}

The winner permutations for multi-thread can be seen in Figure 4.10. It is very interesting that the top based on cycles is very similar to the top in the single thread experiment. The only difference is one neighbour swap of the innermost loops. The top worst case now has a different outermost loop. The top based on L2 misses has an outermost loop order that is not very promising. It is a point on the left graph of Figure 4.9 (left), which tells us that the top based on L2 misses was not indicative for performance this case.

\begin{figure}[h]
\centering
 \includegraphics[trim=1.5cm 0 0 0,width=1.1\textwidth]{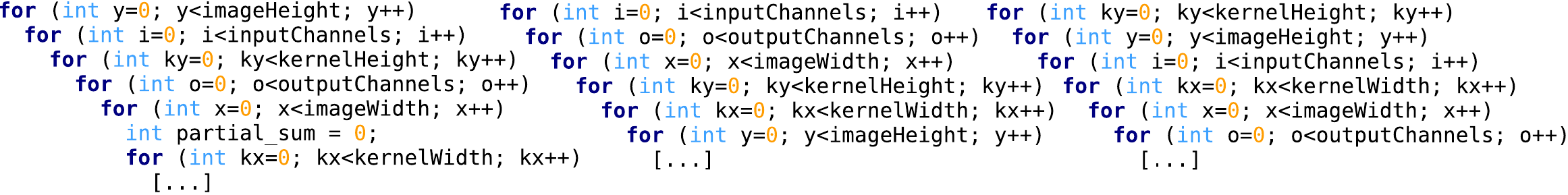}
\caption{The 3 candidates for best loop permutations for multi-thread. From left to right: Top average speedup (0.775002), Top worst-case speedup (0.558273 , average 0.691414), Top speedup based on L2 misses (0.210397)}
\end{figure}

The top permutation based on L2 misses has rank 32 (based on misses) in the single-thread version and therefore the parallelisation is not the reason for producing a small number of misses. It has relatively many L1 misses but few L2 misses. It would be interesting to investigate why. One assumption is that the working set fits well in L2 cache but not in L1 and also there is little reuse between those cycles of the big working set. As we can see, the outermost loop is the output channels which iterates over a relatively big array, which could flush the direct-mapped L1 cache but it fits in L2. Since we are interested in permutations with low L2 misses, we still need to evaluate the resulting permutation candidates. We could select the next top based on L2 misses for many threads.

\chapter{Offline analysis of simulation results on Loop Reordering}

\section{Impact of cache hierarchy}

It is important to investigate whether our analysis is valid for different cache hierarchy configurations. In this way we will be confident that the winner permutations will be valid for different memory hierarchy configurations on Loki, since its reconfigurability is extended on the cache sizes. Of course, if we conclude that cache hierarchy has little impact on top permutations performance, it will also mean that we will not need to manipulate the loop order when searching for optimal L2 cache size or evaluating other optimisations. It is desirable to prove orthogonality between the loop permutation decision and the cache hierarchy because it reduces the design space for other optimisations and generalises the solutions.

The design space for this experiment is the reduced design space of Table 4.3 times three, because I also explore three very different cache hierarchy configurations. The simulated caches combinations are 1) 16KB L1 with 128KB L2, 2) 32KB L1 with 512KB L2 and 3) 64KB L1 with 960KB L2. The number of simulated instructions was 200 millions for every run.

The Figure 5.1 is a parallel coordinates visualisation that shows how the average performance of each individual permutation for all layers varies when changing the cache hierarchy. Each of the 720 lines represents a loop permutation and each parallel axis one cache configuration. The line colouring is based on the hamiltonian index (see Chapter 3). This means that lines of similar colors are permutations that differ only by a couple of neighbour swaps and this is observable in the graph as we can see some kind of clustering of the colors. The main observation is that the top permutations perform almost equally well across all hardware configurations. This is not the case for the non-optimal permutations as they get displaced by a considerable amount across the parallel axes, especially at the first transition from smallest caches.

\begin{figure}[h]
\centering
 \includegraphics[trim=4cm 0 0 0,width=1.1\textwidth]{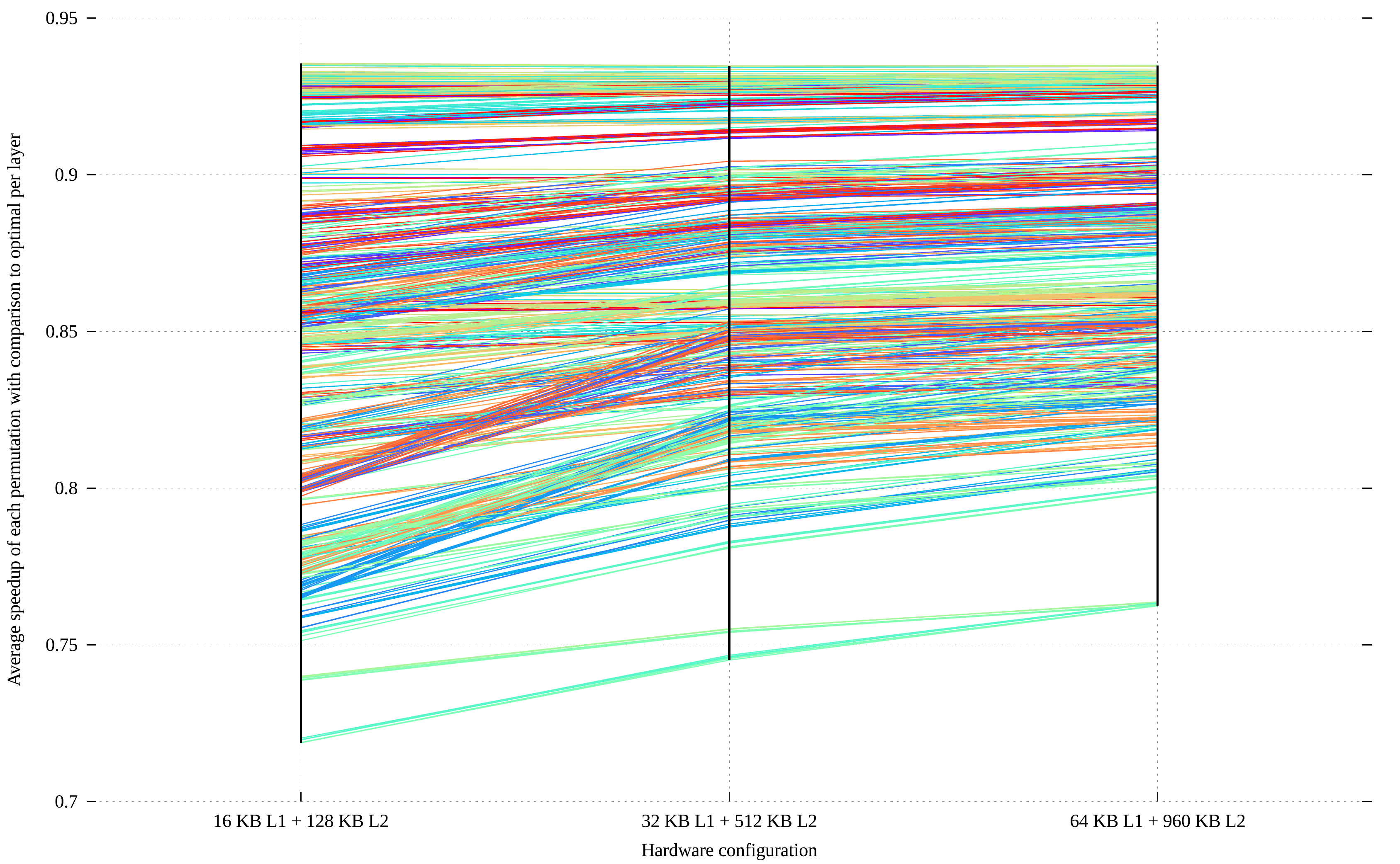}
\caption{Parallel coordinates visualisation for the performance of each of the 720 loop permutation across 3 different cache hierarchy configurations.}
\end{figure}

\section{Impact of multi-threading}

It is also very useful to know if a good loop permutation is good for different numbers of threads. In order to attempt generalisation, we assume that the cache sizes are orthogonal to the loop permutations. The previous section supports this statement to a certain degree. In order to evaluate the impact of threads we set up a similar experiment. The results are based on three data sets, one for 1-thread, one for 4-threads and one for 8-threads. The first and last are already described in the previous chapter. The 4-thread version has the same design space as the 8-thread version, which is described in Table 4.3.

It is also important to note that this result is less representative for Loki in comparison with the previous section because the cache simulator does not differentiate between reads or writes and the safety measures can be more expensive than what is predicted.

In Figure 5.2 we can see how each loop permutation’s average performance changes when increasing the number of threads. As we can see, when we are moving away from the single-thread case, exactly one third forms a group of bad performing permutations. All these permutation are the ones that have either the kernel height or kernel width as the outermost loop, which iterate one or few times and offer little or none exploitable parallelisation. One important observation is that the remaining two thirds of permutations perform fairly similarly when changing the number of threads. However, it is not at the same degree as in the cache sizes impact graph. The most important group of permutations are the top performing ones and their rank seems to change across the different number of threads but by not at a high degree for the two thirds group.

\begin{figure}[h]
\centering
 \includegraphics[trim=3cm 0 0 0,width=1.1\textwidth]{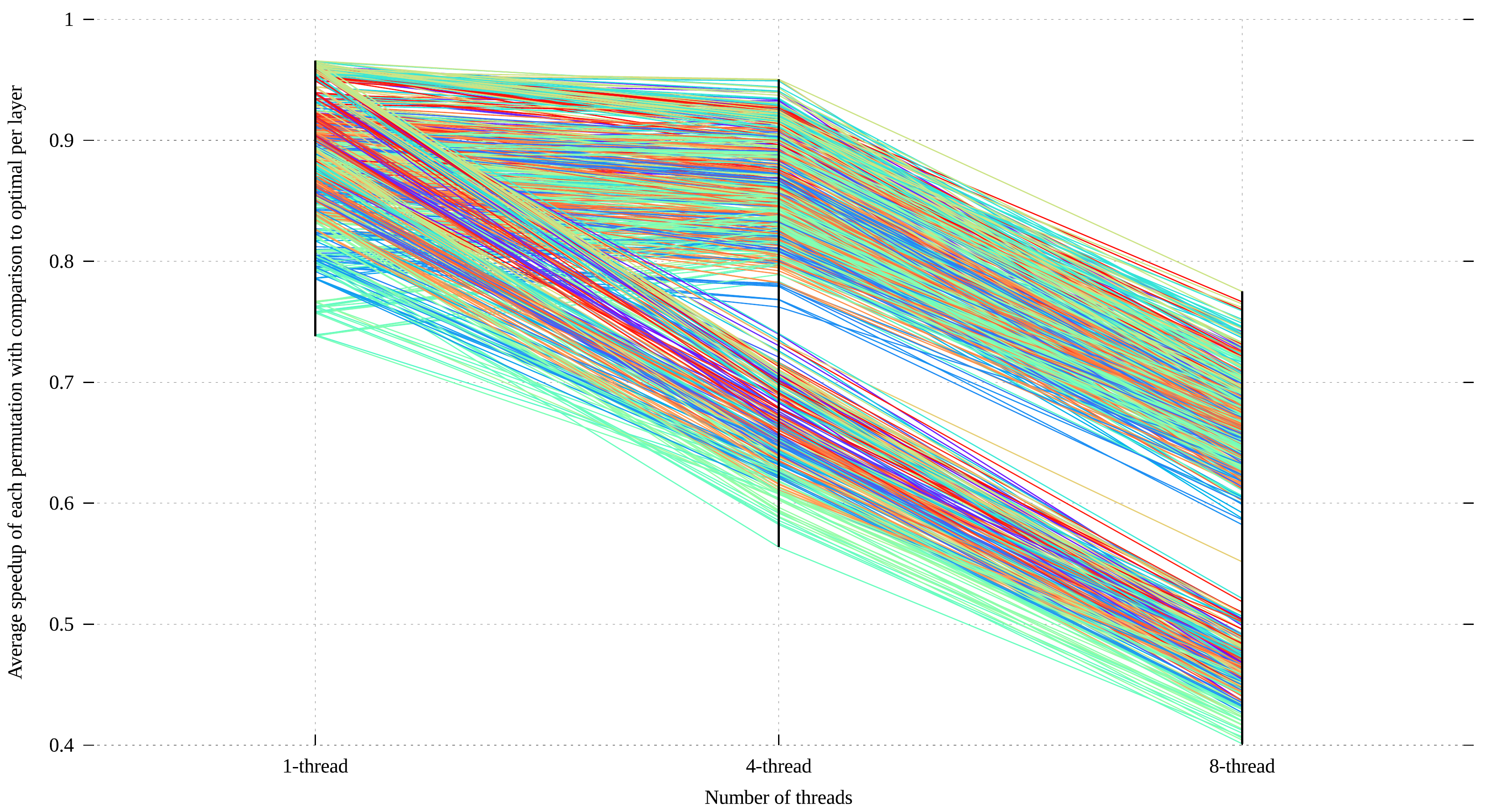}
\caption{Parallel coordinates visualisation for the performance of each of the 720 loop permutation across different number of threads.}
\end{figure}

\section {Dynamic loop reordering}

In this section I do some offline analysis of the results to identify properties that could benefit dynamic loop reordering scenarios. The idea is to create an algorithm that tests a small number of permutations and decides a good performing one for the rest of the execution of the program. This could be used in combination with micro-profiling to eliminate the testing time of the candidate permutations (or tested permutations in general). I evaluate some ideas by using the already calculated results for the layer design space and all permutation results mainly for the 1-thread case.

\subsection {Combinations}

Instead of selecting a single static permutation we could have a small number of permutations where they perform better in separate cases. This selection of permutations can look very different from the top single cases but the should collectively perform equally or better than the top single on average. Ideally, a micro-profiling \cite{5} mechanism could test a very small number of kernels, which are the permutation implementations in our case and select the best performing for the rest of the execution.

In Figure 5.3 we see the equivalent of Figure 4.7 for combinations of 2. Each dot represents the average speedup that would be achieved if we always selected the best out of the permutations in the pair for each layer of the design space (216). As explained before, the speedup range is from 0 to 1 because 1 represents the case where we select the minimum cycles (out of the 720) value that we collected for each layer. Figure 4.7 also could be considered to be for combinations of 1 permutations, Figure 5.3 for combinations of 2 and the optimal value is coming from the combinations of 720.

What Figure 5.3 (left) really tells us is that if we were able to magically select the best permutation out of the top pair for each layer of the design space, we would achieve a 0.99 average speedup over the optimal permutation for each layer; and worst case around 0.83. The biggest benefit of the top pair is better observed in Figure 5.3 (right). The top 1 pair achieves an average theoretical speedup of 0.91 and worst case speedup of 0.68, which is much better than having to chose only from one permutation.

\begin{figure}[h]
\centering
 \includegraphics[trim=1.6cm 0 0 0,width=1.05\textwidth]{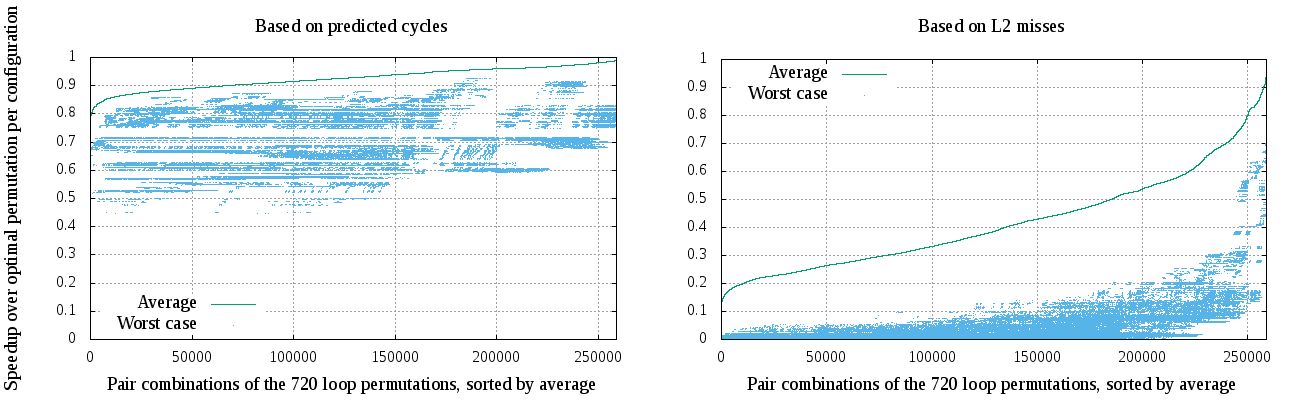}
\caption{Speedup each pair achieves on every layer in comparison with the layer's optimal permutation, on average. On the Left, the average speedup is based on the cycles metric, while on the Right it is based on the L2 misses. When a pair is used instead of a single permutation, only the maximum performance counts when tested for a layer.}
\end{figure}

I have also calculated the respective results for the 8-thread design space and the observations are similar. Again, the benefit is more apparent in the results based on L2 misses. We also could try the same experiments for combinations of more than two permutations.

For the top pairs we can apply a machine learning technique to produce a decision tree. If the decision tree can classify the layer parameter combination with a low true error, then this tree can be used as a simple heuristic to chose between the two permutations of the top pair. If we apply the resulting heuristic on a Loki implementation, it will have no impact on run-time because no profiling is required.

\subsection{Random selection}
 
Testing permutations based on a small random subset would be very straightforward to implement. The question is how big this sample would need to be to have a high probability of finding a good performing permutation. We assume that a good permutation for a single input would perform with speedup of 0.90 over the optimal permutation. In Figure 5.4, there are 216 lines one for each layer, which show the performance of each permutation for each layer. If we take the worst case for speedup equal to 0.90, this layer has only 80 “good” permutations. After applying simple statistics we get that we need 10 random permutations for getting a good permutation for an accuracy of over 68.3\% (one sigma) and 26 for an accuracy of over 95.4\% (two sigma). This result is only for the one-thread results.

\begin{figure}[h]
\centering
 \includegraphics[trim=0 0 0 0,width=0.8\textwidth]{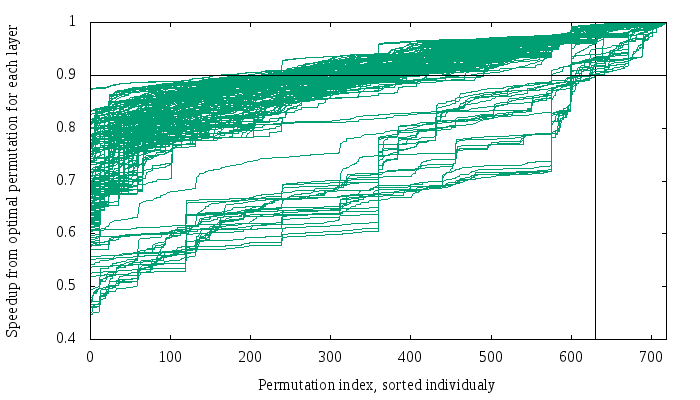}
\caption{Performance of all permutations on all layers, sorted individually. The performance metric is the speedup of the current permutation over the best permutation of the layer. Based on the 1-thread results. }
\end{figure}

While the combinations result sounds more reasonable for dynamic loop reordering, the result of this subsection also suggests a way of reducing the design space when searching for the best permutations for different number of inputs or hardware configurations.

\chapter {Evaluation on Lokisim}

In this chapter I present some results from the lokisim simulator. Lokisim is more accurate than my cache simulator because it models more aspects of the architecture than only the memory hierarchy. Therefore, it is also slower and this is the reason that the full design exploration was not done in this lower-abstraction simulator. This step is to evaluate that the methodology was correct and beneficial for future experiments. I also present the results of some other analyses to show the potential of other optimisations. The algorithm implementations were done by the Loki team.

\section{Performance of top candidates}

In this section I present the results of two of the three candidate permutations of the single-thread case in a small design space of layers. The selected candidates are the top one based on average speedup and the top one based on L2 misses. In Figure 6.1 we can see the set of results along with some other arbitrary permutations. The set of results is limited because of the difficulty that is involved for implementing efficient Loki software at the moment. Additionally, there are some missing data points and the design space is limited because of shortcomings from the experimental state of the Loki toolchain.

\begin{figure}[h]
\centering
 \includegraphics[trim=5cm 0 0 0,width=1.15\textwidth]{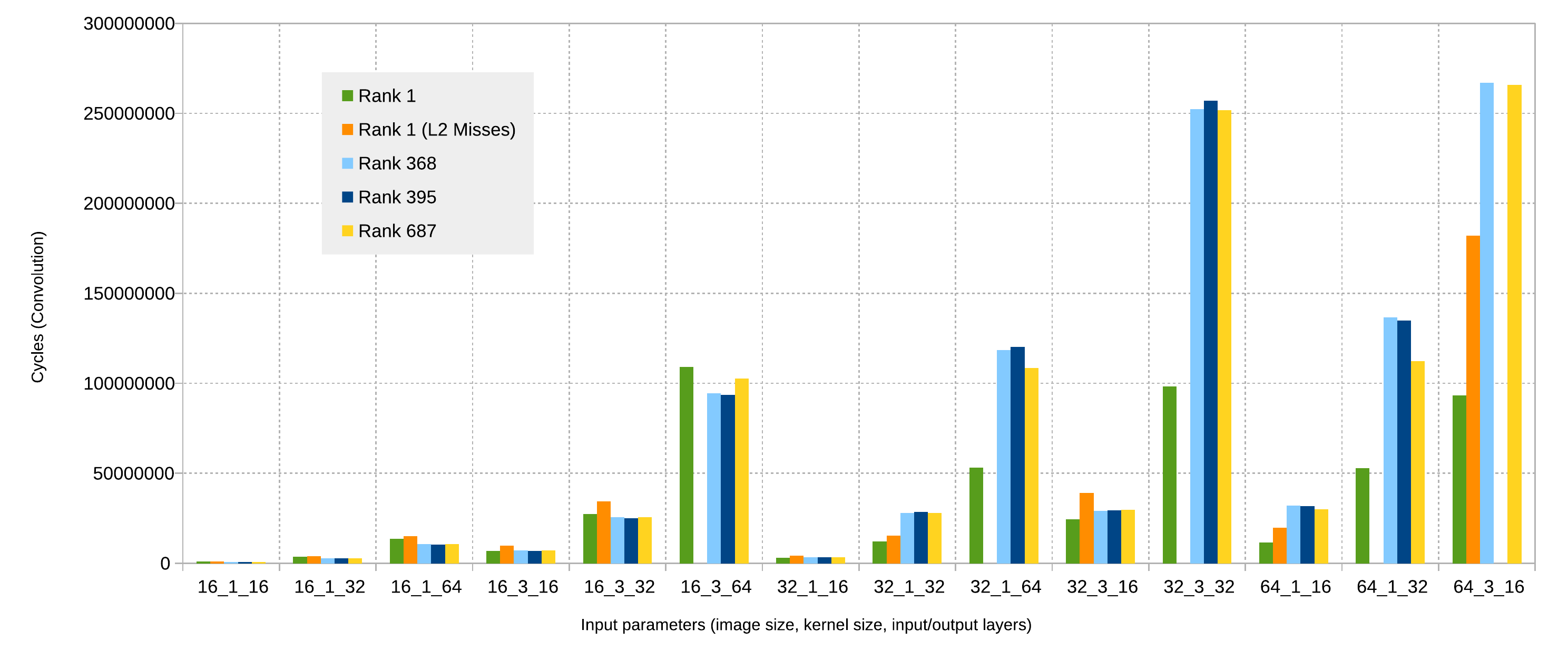}
\caption{Evaluation of 2 candidates for single-thread along with arbitrary permutations in lokisim.}
\end{figure}

As we can see the Rank 1 based on the analysis performs best in comparison with the other selection of permutations in the majority of the layers, especially in the bigger problem-size area (rightmost). We also correctly predicted the two permutations that are shown in blue to be similarly performing according to the rank. The rank 687 performed similarly but not worse than the others. From section 5.1 we saw that the most reliable permutation performance across different memory hierarchy are the top ones. The experiments on Loki had different memory hierarchy that what the analysis was based on and therefore changes in ranks below the top were expected.

The Rank 1 based on L2 misses seems promising, at least for larger layers, but there are many missing data points that would make the argument stronger. The L2 size in the experiment is only 64 KBytes and it could perform even better if we matched the L2 size of the previous analysis.

\section{Comparison with sparsity algorithms}

I also present a small analysis on existing Loki convolution implementations to show the impact of the sparsity measures and the impact of loop order. These results are based on implementations which were manually optimised and with different cache partitioning layouts. It would be interesting to see how my candidates compare to this result, but it will be a future work.  

In Figure 6.2 we see the results for three configurations and different inputs on the same input parameters. The inputs are synthetic images of random data with specified weight and activation density. As we can see from the graph, the two dense implementations are completely insensitive to the input sparsity characteristics. The sparse algorithm, shown in blue, uses one core that searches for non-zero data and the remaining cores of the tile are doing the computation on request. In this way it saves a lot of time when the image or the weights array has a lot of zeros.

\begin{figure}[h]
\centering
 \includegraphics[trim=1.65cm 0 0 0,width=1.02\textwidth]{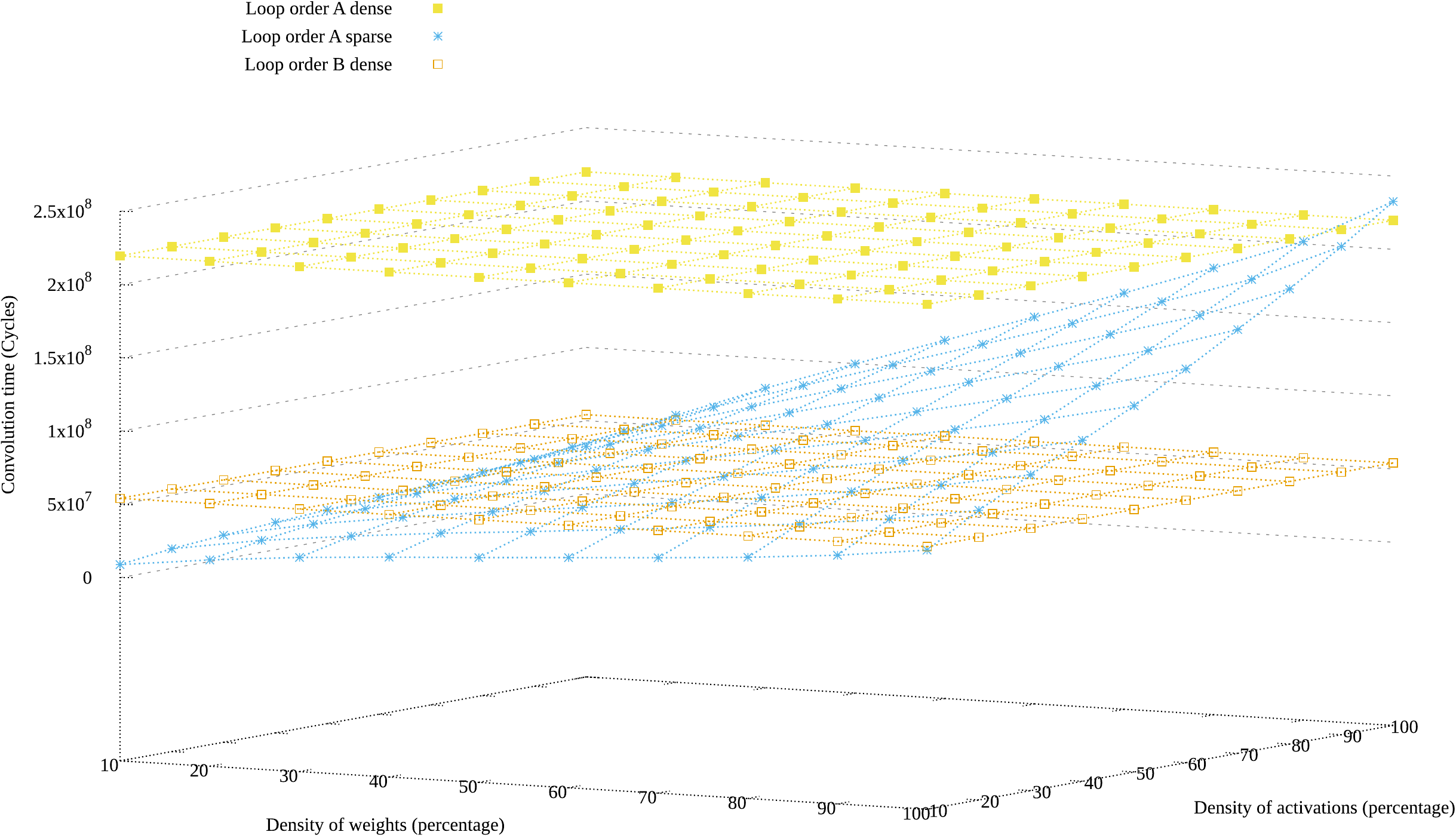}
\caption{Performance in cycles for convolution by using 3 different algorithms and different inputs. Parameters: Image size 25x25, kernel size 3x3, 128 input and output channels. Architecture configuration: 1 tile for computation and 1 tile for L2 cache (64 Kbytes)}
\end{figure}

We can also see the impact of the loop order choice in these highly optimised implementations. The loop order B is over 4 times faster than the loop order A. This is an important finding because we did not observe this kind of speedup in my small-design space loop order analysis. It tells us that loop order is more important on real hardware because there are other bottlenecks that could make low-locality access patterns perform even worse. 

The dense version of the A loop order is better than the sparse version only for the input image and weights of 100\% sparsity each, which is an unrealistic case. If we compare the sparse version of A with the dense of B the answer is more complicated. The sparse implementation wins at the low density cases but B performs better in the majority of runs. Therefore, in order to decide which of the two would be the best case we would need to study the average case for the density of activations and weights. Ideally, we would also find a best performing loop order that will also be friendly for sparse implementations.

\section{Swapping Tiles for L2 cache}

This experiment explores the impact of swapping computation tiles for a bigger unified L2 cache. The simulated hardware instance has 16 tiles in total. I also explore the potential of selecting the optimal number of tiles for L2 cache and computation.
 
As a first experiment I tried all the different combinations of the number of tiles for each purpose for a single layer. In Figure 6.3 we see how the performance is affected by the tiles configuration. For this example the optimal number of compute tiles is 10 and the number of L2 tiles are the remaining 6. One observation is that the more computation tiles exist, the more useful a bigger L2 cache is. On the diagonal line we have full utilisation of the tiles and since we attempt full utilisation, the number of tiles for L2 is equal to the difference of 16 and the number of computation tiles. We can express these configurations with only one of the two parameters.

\begin{figure}[h]
\centering
 \includegraphics[trim=1.65cm 0 0 0,width=1.03\textwidth]{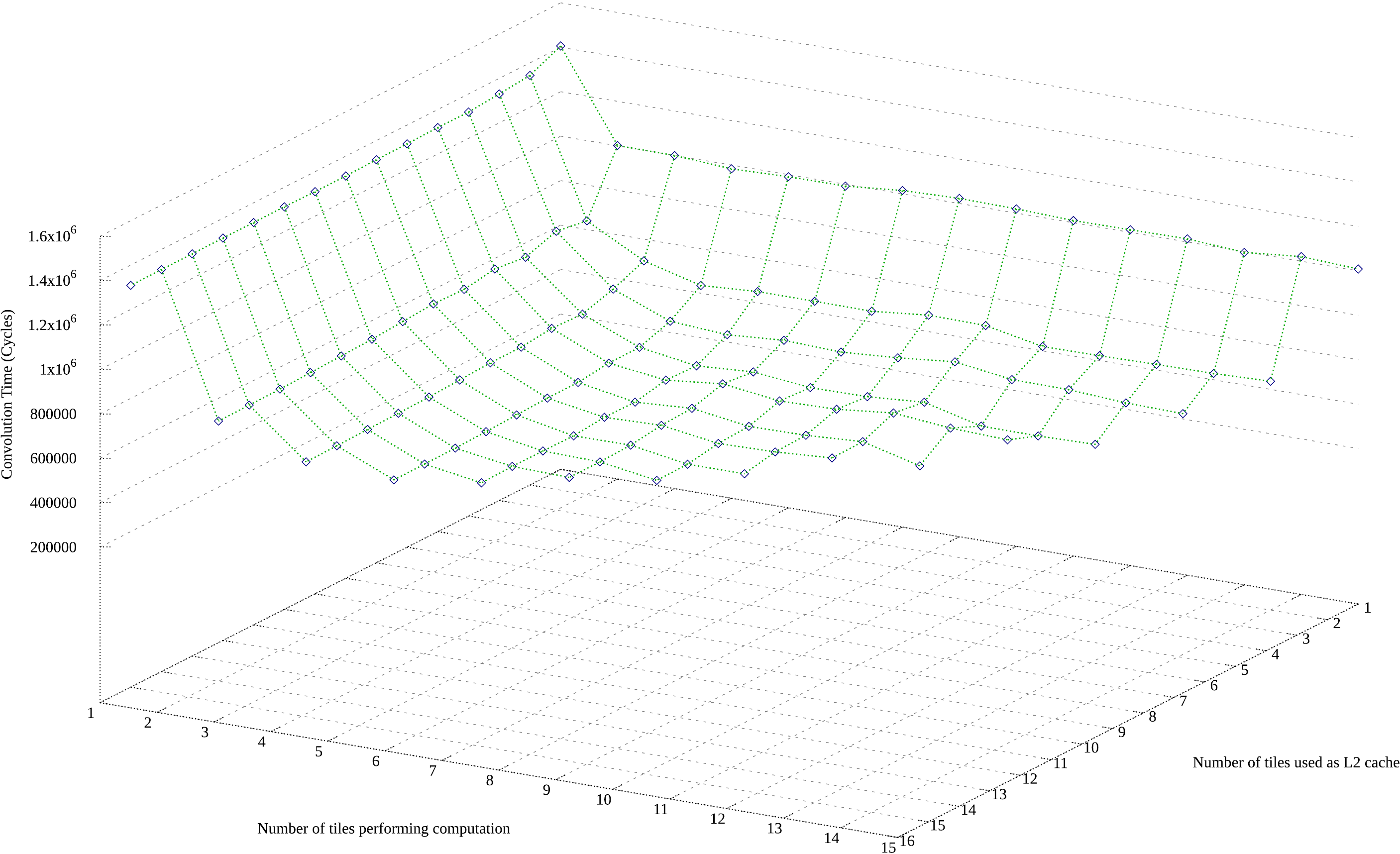}
\caption{Performance in cycles for all possible combinations of the number of tiles for computation and the number of tiles for L2}
\end{figure}

I also explore the potential for dynamically selecting the optimal configuration for each layer. In the following experimentation I used a small design space of layers consisting of different number of inputs and output channels. I ran all the 15 value combinations for full utilisation for each of the layers and found the best performing overall. Then I compared it to the result of the optimal per layer configuration to see what could be achieved by making this decision on the fly. The best overall tile configuration in this case was 8 tiles for computation and 8 tiles for L2 cache. Figure 6.4 shows that there is a common winner among the combinations of bigger number of input channels and the average speedup when selecting the optimal tile configuration would only be 1.5\% and at most around 12\%. This speedup may not justify the introduction of a dynamic tile configuration for convolution but it would be useful to verify this observation with other parameter combinations as well.

\begin{figure}[h]
\centering
 \includegraphics[trim=2.45cm 0 0 0,width=1.015\textwidth]{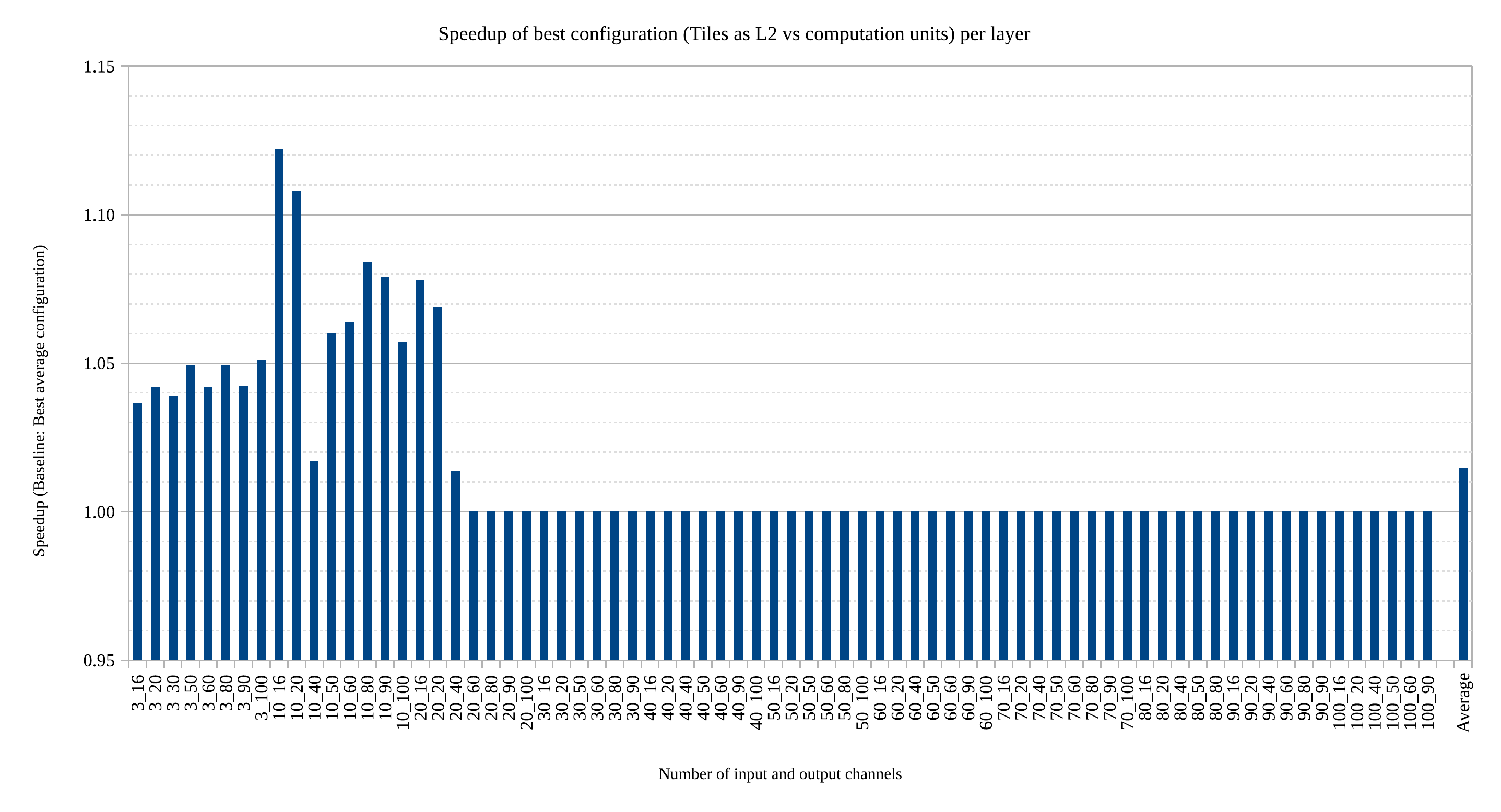}
\caption{Speedup of the best tile configuration over the best average (8 tiles for calculation, 8 for L2) for a set of layers with different number of input channels and output channels. }
\end{figure}

\section{Motivation for adaptive algorithms}

In the last section we saw that there would not be much need for dynamic swapping of tiles for L2 for the respective layer parameter ranges. However, it would be still useful to see weather we could have an accurate prediction mechanism that would make correct choices under different configurations. In Figure 6.5 we can see how the recent IPC could predict the total execution time for 15 different tile configurations. The drawn lines are smoothed for better demonstration. As we can see, after the initialisation phase the recent IPC is a very good indicator for the overall performance. This is because it remains steady throughout the convolution execution due to the simple access patterns and algorithm that convolution has. IPC could be used to profile small sections of different versions of the code or other configurations to make correct choices in a small amount of time and therefore little overhead.

\begin{figure}[h]
\centering
 \includegraphics[trim=1.3cm 0 0 0,width=1.019\textwidth]{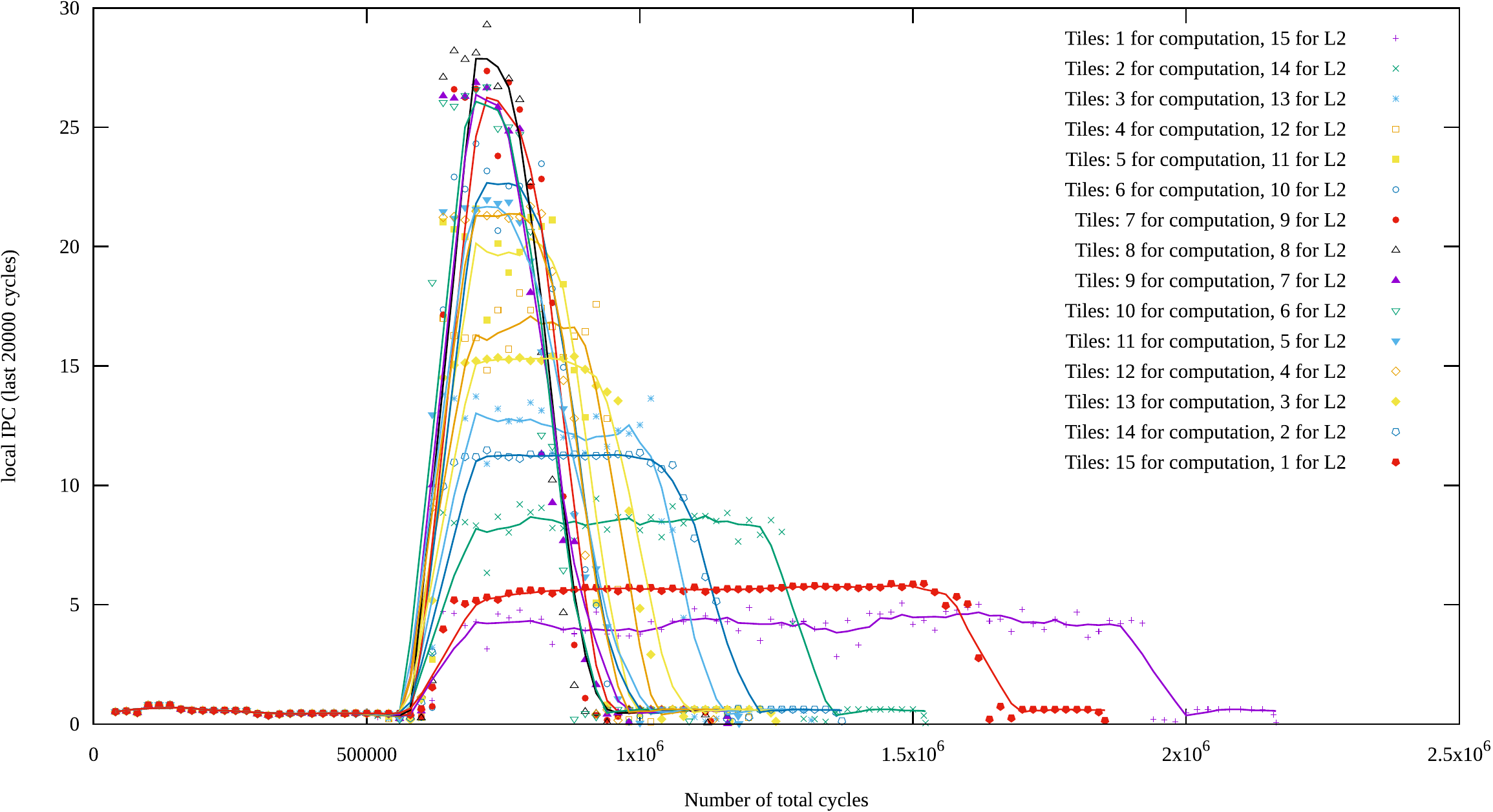}
\caption{Recent IPC during the full convolution execution for 15 different tile configurations. }
\end{figure}

\chapter{Conclusions}

\section{Summary}

In this project I explored a design space of optimisations for Loki by using two simulation frameworks of different levels of abstraction.

From the loop order analysis I produced some candidate loop permutations for evaluation on lokisim by performing an exhaustive search for the best loop permutations in a fast cache simulator. This kind of loop reordering analysis could also be applied in similar problems where there are a lot of independent nested loops.

From the offline analysis of the simulator data, I explored some properties of loop orders that could be used for dynamic loop ordering or faster design space exploration. I showed that instead of having a single top loop permutation we could select the top combination of N loop permutations which can collectively perform near optimally on average. I also explored the prospects when selecting a limited random sample of permutations for finding a good performing permutation.

I also expanded the design space for 3 different hypothetical architectures and 3 levels of parallelism. This demonstrated that the top orders will still perform optimally when changing cache hierarchy and also when changing the number of threads from one to eight. It is usually challenging to prove orthogonality when the design space is limited by the total simulation time but in this case a variety of results from different hardware and software configurations, including on lokisim, validated the stability of the top loop orders. The validation on lokisim also proved the correctness or usefulness of the methodology.

In the evaluation on Lokisim chapter I also explored the potential of other optimisations such as dynamically swapping tiles for L2 cache and the sparsity algorithms. We concluded that dynamically changing tile configurations would have limited speedup (around 10\% at maximum) over an optimal tile configuration. In the same chapter, however, I showed that adaptiveness to performance would be very applicable in a convolution algorithm because even in different parallel versions with different sorts of bottlenecks, the recent IPC predicts accurately the total execution time. This finding also validated the methodology because in the simulations we were stopping the execution early for wider design exploration.

\section{Future Work}

There are many ways this analysis can be expanded. The design space of each experiment was limited to save simulation time and it could be useful to repeat the experiments for different number of threads and layer configurations (input parameters). There are also many other optimisations that can be applied and they all contribute to a wide design space for future exploration.

In section 5.2 I showed that there is a correlation between the top loop permutations when changing the number of threads. This was limited for the single-core, four-core and eight-core configurations. It is a positive result for showing that top loop permutations will scale for bigger number of threads but we would like to see if this continues for more threads. My cache simulator already supports multi-tile configurations and it would be very useful to know if the section of the best permutations remains the same for multi-tile configurations.

Regarding access pattern manipulation, our focus was on the loop order. However, since the block access pattern is different from the reference access pattern, the way each multi-dimensional array is represented in memory could have important impact on performance. Therefore it could be one other optimisation technique that would be interesting to explore. The exploration could be combined with the permutations analysis. One hypothesis could be that the permutations would have different results but there would be equivalent permutations with similar loop performance results. 

One other way of improving cache performance is loop tiling \cite{28}. Loop tiling is a loop transformation optimisation where a a nested loop is split into smaller blocks to minimise the working set and produce less capacity misses. This transformation would also be a nice extension to this analysis because each of the two optimisations seem very dependent to each other's decision and the combined design space could lead to more efficient solutions

The practical evaluation of a form of dynamic loop reordering would probably benefit from this analysis. A current approach for profiling and selecting the best algorithm is micro-profiling \cite{5}. Microprofiling is proven to increase the performance of massively-parallel software on GPUs by selecting the optimal kernel to be used throughout the whole execution of the program and making changes to the algorithm on the fly. In the neural networks case I evaluate approaches similar to microprofiling for further optimisation during runtime. Some algorithms have already been discussed for applications on micro-profiling or heuristics to select the best performing loop permutation at run-time and could be easily evaluated on specific simulators and hardware as a test platform. For evaluation of micro-profiling on conventional architectures, the PAPI framework \cite{29} could be used to read the performance counter values and compare and switch permutations during runtime.

There are also some other ideas that I would like to explore for potential in dynamic loop reordering. The first is to apply parameter optimisation using machine learning methods for exploring efficiently permutations on runtime. There would be a single parameter, the hamiltonian index, whose spatial localities could save time in searching. Another idea is to search by using Breadth-First Search on the permutations graph based on neighbour swap (see Figure 4.2). The latter would probably perform better because the graph contains much more locality information than a linear function, although it could be more difficult and costly to implement graph traversal algorithms in a micro-profiling environment.

As we have shown from experimentation on lokisim, measures such as IPC remain steady during the parallel execution of the convolution. This means that a micro-profiling mechanism could perform well with small amounts of sampling for this application. The results are very promising for adaptive algorithms on convolution because we found a simple metric for quick comparison of implementations.

\appendix
\singlespacing


\addcontentsline{toc}{chapter}{Bibliography}

\begin{thebibliography}{9}

\bibitem{1} 
Bates, D., Chadwick, A., \& Mullins, R.  
\textit{Configurable memory systems for embedded many-core processors.}. 
2016.
 
\bibitem{2} 
Bates, D., Bradbury, A., Koltes, A., \& Mullins, R. 
\textit{Exploiting tightly-coupled cores. Journal of Signal } 
Processing Systems, 80 (1), 103-120., 2014
 
 


\bibitem{5}
Li-Wen Chang, Hee-Seok Kim, Wen-mei Hwu 
\textit{DySel: Lightweight Dynamic Selection for Kernel-based Data-parallel Programming Model}
ASPLOS '16

\bibitem{6}
Ovtcharov, K., Ruwase, O., Kim, J.Y., Fowers, J., Strauss, K. and Chung, E.S., . 
\textit{Accelerating deep convolutional neural networks using specialized hardware.}
Microsoft Research Whitepaper, 2(11). 2015

\bibitem{7}
Gaj, K. and Chodowiec, P., 2009. In Cryptographic engineering (pp. 235-294). 
\textit{ FPGA and ASIC implementations of AES.}
Springer US.

\bibitem{8}
Kottolli, A., 2006. 
\textit{The economics of structured-and standard-cell-ASIC designs. }
Electronic News, 16, p.2006.

\bibitem{9}
Park, Y., Park, J.J.K. and Mahlke, S.   
\textit{Efficient performance scaling of future cgras for mobile applications.}
In Field-Programmable Technology (FPT), 2012 International Conference on (pp. 335-342). IEEE. 2012, December.

\bibitem{10}
Bates, Daniel, Alex Bradbury, Andreas Koltes, and Robert Mullins. 
\textit{Spatial computation on a homogeneous, many-core architecture }
(2014).

\bibitem{11}
Ramchoun, H., Amine, M., Idrissi, J., Ghanou, Y., \& Ettaouil, M. 
\textit{ Multilayer Perceptron: Architecture Optimization and Training. International}
 Journal of Interactive Multimedia and Artificial Intelligence, 4(Special Issue on Artificial Intelligence Underpinning). (2016)

\bibitem{12}
Parekh, R., Yang, J., \& Honavar, V.   
\textit{Constructive neural-network learning algorithms for pattern classification.}
IEEE Transactions on neural networks, 11(2), 436-451. (2000).

\bibitem{13}
Krizhevsky, A., Sutskever, I. and Hinton, G.E., 2012. 
\textit{ Imagenet classification with deep convolutional neural networks.}
In Advances in neural information processing systems (pp. 1097-1105).

\bibitem{14}
Iandola, F.N., Han, S., Moskewicz, M.W., Ashraf, K., Dally, W.J. and Keutzer, K., 2016.
\textit{SqueezeNet: AlexNet-level accuracy with 50x fewer parameters and< 0.5 MB model size}
arXiv preprint arXiv:1602.07360.

\bibitem{15}
Ballester, P. and Araujo, R.M., 2016, February. 
\textit{On the performance of GoogLeNet and AlexNet applied to sketches. }
In Thirtieth AAAI Conference on Artificial Intelligence.

\bibitem{16}
Smith, S.W., 1997. 
\textit{Chapter 18 – FFT Convolution}
 The scientist and engineer's guide to digital signal processing.

\bibitem{17}
Luk, C.K., Cohn, R., Muth, R., Patil, H., Klauser, A., Lowney, G., Wallace, S., Reddi, V.J. and Hazelwood, K., 2005, June. 
\textit{Pin: building customized program analysis tools with dynamic instrumentation.}
 In Acm sigplan notices (Vol. 40, No. 6, pp. 190-200). ACM.

\bibitem{18}
Intel Pin 2.14 User Guide
\textit{Memory Reference Trace (Instruction Instrumentation)}
\url{https://software.intel.com/sites/landingpage/pintool/docs/67254/Pin/html/index.html#MAddressTrace} Retrieved on 11 June 2017


\bibitem{19}
Brendan Gregg and Jim Mauro
\textit{DTrace: Dynamic Tracing in Oracle Solaris, Mac OS X and FreeBSD}
Prentice Hall Press, Upper Saddle River, NJ, USA, April 2011.

\bibitem{20}
Wagner, P., Wild, T. and Herkersdorf, A., 2016. 
\textit{Improving SoC Insight Through On-Chip Diagnosis.}
arXiv preprint arXiv:1607.04549.

\bibitem{21}
Johnson, S.M., 1963.  
\textit{Generation of permutations by adjacent transposition.}
Mathematics of computation, 17(83), pp.282-285.

\bibitem{22}
Cho, S. and Jin, L., 2006, December. 
\textit{Managing distributed, shared L2 caches through OS-level page allocation. }
In Proceedings of the 39th Annual IEEE/ACM International Symposium on Microarchitecture (pp. 455-468). IEEE Computer Society.

\bibitem{23}
Jaleel, A., Najaf-Abadi, H.H., Subramaniam, S., Steely, S.C. and Emer, J., 2012, March.
\textit{CRUISE: cache replacement and utility-aware scheduling. }
 In ACM SIGARCH Computer Architecture News (Vol. 40, No. 1, pp. 249-260). ACM.

\bibitem{24}
Chatterjee, N., Shevgoor, M., Balasubramonian, R., Davis, A., Fang, Z., Illikkal, R. and Iyer, R., 2012, December.
\textit{ Leveraging heterogeneity in DRAM main memories to accelerate critical word access. }
In Microarchitecture (MICRO), 2012 45th Annual IEEE/ACM International Symposium on (pp. 13-24). IEEE.

\bibitem{25}
Joseph Redmond 
\textit{ TinyDarknet}
\url{https://pjreddie.com/darknet/tiny-darknet/} Retrieved on 11 June 2017

\bibitem{26}
Perelman, E., Hamerly, G., Van Biesbrouck, M., Sherwood, T. and Calder, B., 2003, June.  
\textit{Using SimPoint for accurate and efficient simulation.}
In ACM SIGMETRICS Performance Evaluation Review (Vol. 31, No. 1, pp. 318-319). ACM.

\bibitem{27}
Sugumar, R.A. and Abraham, S.G., 1993. 
\textit{Efficient simulation of caches under optimal replacement with applications to miss characterization }
(Vol. 21, No. 1, pp. 24-35). ACM.

\bibitem{28}
Abella, J., González, A., Llosa, J. and Vera, X., 2002. 
\textit{Near-optimal loop tiling by means of cache miss equations and genetic algorithms. }
In Parallel Processing Workshops, 2002. Proceedings. International Conference on (pp. 568-577). IEEE.

\bibitem{29}
Terpstra, D., Jagode, H., You, H. and Dongarra, J., 2010. 
\textit{Collecting performance data with PAPI-C.}
In Tools for High Performance Computing 2009 (pp. 157-173). Springer Berlin Heidelberg.


\bibitem{30}
Allen, J.R. and Kennedy, K., 1984, June. 
\textit{Automatic loop interchange.}
 In Acm Sigplan Notices (Vol. 19, No. 6, pp. 233-246). ACM.

\bibitem{31}
Cain, H.W., Michael, M.M., Frey, B., May, C., Williams, D. and Le, H., 2013.  
\textit{Robust architectural support for transactional memory in the power architecture.}
ACM SIGARCH Computer Architecture News, 41(3), pp.225-236.

\bibitem{32}
Pradel, M. and Gross, T.R., 2012. 
\textit{Fully automatic and precise detection of thread safety violations.}
 ACM SIGPLAN Notices, 47(6), pp.521-530.

 
 
 
\end{thebibliography}
\end{document}